\documentclass[aps,prd,twocolumn,amsmath,amssymb,superscriptaddress,nofootinbib]{revtex4-2}

\usepackage{graphicx}
\usepackage{xcolor}
\usepackage{soul}

\newcommand{\orcid}[1]{\href{https://orcid.org/#1}{\includegraphics[width=9pt]{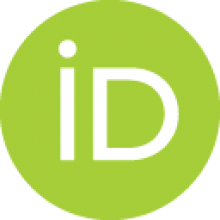}}}

\usepackage{aas_macros}
\usepackage[colorlinks,citecolor=blue]{hyperref}

\begin{document}

\title{Viscous hydrodynamic evolution of neutron star merger accretion disks: a code comparison }

\author{Rodrigo Fern\'andez\orcid{0000-0003-4619-339X}}
\email[]{rafernan@ualberta.ca}
\affiliation{Department of Physics, University of Alberta, Edmonton, AB T6G 2E1, Canada}

\author{Oliver Just\orcid{0000-0002-3126-9913}}
\affiliation{GSI Helmholtzzentrum {f\"ur} Schwerionenforschung, Planckstra{\ss}e 1, D-64291 Darmstadt, Germany}
\affiliation{Astrophysical Big Bang Laboratory, RIKEN Cluster for Pioneering Research, 2-1 Hirosawa, Wako, Saitama 351-0198, Japan}

\author{Zewei Xiong\orcid{0000-0002-2385-6771}}
\affiliation{GSI Helmholtzzentrum {f\"ur} Schwerionenforschung,
  Planckstra{\ss}e 1, D-64291 Darmstadt, Germany}

\author{Gabriel Mart\'inez-Pinedo\orcid{0000-0002-3825-0131}}
\affiliation{GSI Helmholtzzentrum {f\"ur} Schwerionenforschung,
  Planckstra{\ss}e 1, D-64291 Darmstadt, Germany} 
\affiliation{Institut f{\"u}r Kernphysik (Theoriezentrum),
    Fachbereich Physik, Technische Universit{\"a}t Darmstadt,
    Schlossgartenstra{\ss}e 2, D-64289 Darmstadt, Germany} 

\date{\today}

\begin{abstract}
The accretion disk formed after a neutron star merger is an important
contributor to the total ejecta from the merger, and hence to the kilonova and the
$r$-process yields of each event. Axisymmetric viscous hydrodynamic
simulations of these disks can capture thermal mass ejection due to neutrino absorption and
in the advective phase---after neutrino
cooling has subsided---and are thus likely to provide 
a lower-limit to the total disk ejecta
relative to MHD evolution. Here we present a comparison between
two viscous hydrodynamic codes that have been used extensively on
this problem over the past decade: ALCAR and FLASH. 
We choose a representative setup with a black hole at the center, and 
vary the treatment of viscosity and neutrino transport. We find good
overall agreement ($\sim 10\%$ level) in most quantities. The average
outflow velocity is sensitive to the treatment of the nuclear binding energy
of heavy nuclei, showing a larger variation than other quantities. We post-process trajectories 
from both codes with the same nuclear network, and explore
the effects of code differences on nucleosynthesis yields, heating rates, and kilonova
light curves. For the latter, we also assess the effect of varying the number of
tracer particles in reconstructing the spatial abundance distribution for kilonova light curve production.
\end{abstract}

\maketitle

\section{Introduction \label{s:intro}}

Production of chemical elements heavier than iron in the universe via the rapid neutron
capture process ($r$-process) has thus far been established observationally
for neutron star (NS) mergers through the kilonova associated with GW170817 
(e.g., \cite{ligo_gw170817_multi-messenger,drout_2017}).
The accretion disk formed during the merger is a significant or even dominant
contributor to the ejecta---depending on binary parameters---launching outflows
on timescales ranging from a few ms to several seconds after the merger 
(e.g., \cite{FM16,baiotti_BinaryNeutronStar_2017,radice_DynamicsBinaryNeutron_2020,janka_bauswein_2022_review}).

Multiple processes can lead to mass ejection from the disk: dissipation of
magnetorotational turbulence, nuclear recombination, neutrino absorption,
and magnetic stresses if a large-scale magnetic field is present at disk
formation or generated via dynamo action (e.g., \cite{siegel_2018,Just2022_Yeq}). 
Neutrino cooling is important in
all disks with initial masses 
$\gtrsim 10^{-3}M_\odot$ (e.g., \cite{popham1999,DiMatteo+02,setiawan2004,Chen&Beloborodov07,de_siegel_2021}), 
but it subsides on a timescale
of several $\sim 100$\,ms in disks around black holes (BHs) due to the drop in temperature and density associated with
accretion. The absence of cooling leads to
ejection driven by viscous heating and nuclear recombination \cite{Metzger+09a}.
When a NS is present, energy deposition by neutrino absorption can also make a significant contribution
to driving the outflow (e.g., \cite{MF14, Just2015a, fujibayashi_2018, fujibayashi_2020_ns}).

The magnetic field strength and geometry at disk formation determines
the importance of prompt mass ejection due to magnetic stresses and
the possible emergence of a jet (e.g., \cite{christie2019}). These magnetic 
properties are currently an active
area of research. For BH central objects, the only ab-initio study thus far
\cite{hayashi_2022a,hayashi_2022b} indicates that large scale field formation is not ubiquitous, with
the corresponding absence of  prompt ($\sim $\,ms) mass ejection via magnetic stresses.  
Thus, in the case of BH central objects, thermal mass ejection due to the drop 
in neutrino cooling is the only
outflow channel established as robust thus far.

Long-term viscous hydrodynamic models of the disk outflow have been carried
out for a decade now, and have led to most of our current understanding
of the disk ejecta 
\cite{FM13,MF14,Just2015a,FKMQ14,FQSKR-15,fernandez_2017,fujibayashi_2018,fahlman_2018,fujibayashi2020,Fernandez2020BHNS,metzger_2021_late-time,Just2022_Yeq,Just2022_FFI,fujibayashi_2022,fernandez_2022_FFI,haddadi_2023}. 
For BH remnants, these simulations are able to capture thermal
mass ejection to good approximation, as demonstrated by detailed comparison with
GRMHD simulations \cite{F19_grmhd}. Viscous hydrodynamic simulations thus provide
a good estimate for the lower limit to the mass ejection from post-merger
accretion disks (assuming that magnetic effects can only enhance it).

Despite awareness of broad agreement between groups carrying out
viscous hydrodynamic simulations of NS merger disks, a quantitative
code comparison has never been done. Experience from the
core-collapse supernova modeling community shows that
code comparisons help estimating the uncertainties of theoretical predictions
and they can provide valuable insight into the physics of the system, by
unfolding sensitivities with respect to specific assumptions and approximations
adopted by individual codes or models
\cite{messer_1998,yamada_1999,liebendoerfer_2005,richers_2017,just_2018,oconnor_2018,cabezon_2018,varma_2021}.

Here we carry out a quantitative code comparison study between
the viscous hydrodynamic setups of Just and collaborators (based on the
ALCAR code) and Fern\'andez and collaborators (based on a modified version of the FLASH code).
Both setups have been used extensively over the past decade, and model
viscous angular momentum transport, the BH pseudo-Newtonian
potential, and the equation of state in a similar manner. The implementations differ
primarily in the neutrino transport method employed
(multi-group 2-moment [M1] for ALCAR, gray leakage + absorption for FLASH). 
For the comparison, we choose an initial condition with
the same physical parameters, and vary the treatment of viscosity 
as well as the number of neutrino species and neutrino production processes considered. Each
code employs production settings used previously in publications. Thus, while most
of the settings are chosen to be the same between both codes, 
not all numerical details are exactly identical for this comparison study, such as
the computational grid or the implementation of the equation of state.
We study the role of the additional binding energy gained by the formation of heavy nuclei
beyond alpha particles, which has been neglected in some accretion disk models, 
and can have a non-negligible impact on the outflow velocity.
We also generate tracer particles and perform post-processing nucleosynthesis calculations
to assess the effects of code differences on $r$-process abundances, heating rates, and kilonova lightcurves.
For the latter, we also explore how changing the number of particles included
influences the light curves through the spatial distribution of lanthanides
and actinides. 

The structure of this paper is the following. Section~\ref{s:methods} describes
the codes used, approximations to the physics made, and the models evolved.
Section~\ref{s:results} presents our results and analysis, followed by a
Summary and Discussion in Section~\ref{s:discussion}. The Appendix presents
the equations used to determine the composition and internal energy assuming a mixture of neutrons, 
protons, alpha particles, and a representative heavy nucleus in nuclear statistical equilibrium (NSE).

\section{Methods \label{s:methods}}

\subsection{Codes and physics included \label{s:codes}}

\subsubsection{ALCAR \label{s:alcar}}

The ALCAR code \cite{Just2015b}, which is based on the magnetohydrodynamics code AENUS \cite{Obergaulinger2008a}, evolves the viscous hydrodynamics equations along with conservation equations for the 0th and 1st angular moments of the neutrino intensity (energy- and flux-density, respectively) on an axisymmetric spherical-coordinate mesh using finite-volume, high-order shock-capturing methods. ALCAR offers both a Newtonian and special relativistic framework, as well as various schemes for the time-integration, spatial reconstruction, and Riemann solver. Here we adopt the Newtonian version, a 2nd-order Runge-Kutta integrator, the PPM$_5$ scheme of \cite{Mignone2014a}, and the Harten-Lax-van Leer Riemann solver, respectively. Gravity is treated using the pseudo-Newtonian Artemova-Novikov potential \citep{artemova1996}. The equation of state (which was originally implemented in Ref.~\cite{Janka1996b}) assumes a Boltzmann gas of four baryonic species (neutrons, protons, helium, and $^{54}$Mn) in nuclear statistical equilibrium (NSE), a Fermi-gas of electrons and positrons, and a thermal bath of photons. The radial domain of $r\in [10^6\,\mathrm{cm}, 4\times 10^{11}\,\mathrm{cm}]$ is discretized by 576 logarithmically spaced zones, and the polar-angle domain, $\theta\in [0,\pi]$, is sampled by 160 uniform zones.

The ALCAR models presented in this study adopt the same setup as in \cite{Just2015a}, except that here we a) include heavy-lepton neutrinos\footnote{However, $\mu/\tau$ neutrinos only play a subdominant role in BH-disks, at least in the case when neutrino oscillations are ignored (e.g.~\cite{Just2022_FFI}).}; b) evolve the entropy at radii beyond 1000\,km (instead of evolving the sum of kinetic and thermal energy everywhere) in order to prevent numerical artefacts in the expanding ejecta once the thermal energy becomes subdominant; c) start from slightly different initial conditions in the disk (constant entropy instead of polytropic relation with fixed maximum density); d) use online tracers for the ejecta analysis (instead of post-process tracers constructed via backward integration from the written output data). Despite these differences, the results for the ALCAR models discussed here are well in agreement with those in \cite{Just2015a}.

The neutrino transport adopts the M1 approximation, meaning that all higher angular moments (e.g. the Eddington tensor) appearing in the moment equations are expressed as local functions of the evolved moments using a closure relation. We adopt the closure by \cite{Minerbo1978} (in the same form as in Ref.~\cite{Just2015b}). We discretize energy space of neutrinos using 10 energy bins logarithmically spaced between 0 and 80 MeV and evolve the two-moment system for each energy bin. We take into account velocity-dependent terms up to first order in $v/c$ following previous disk studies (see, e.g., \cite{Just2015a,Just2022_Yeq}). The transport follows the evolution of three neutrino species, $\nu_e$, $\bar\nu_e$, and $\nu_x$ (with $\nu_x$ representing the four heavy-lepton neutrinos), which interact with free nucleons via emission and absorption (only $\nu_e$ and $\bar\nu_e$) as well as iso-energetic scattering, with rates taken from \cite{Bruenn1985} and augmented by weak-magnetism corrections \citep{Horowitz2002a}. The production of heavy-lepton neutrinos proceeds through $e^\pm$ annihilation \citep{Pons1998} and Bremsstrahlung \citep{Hannestad1998}, while for the corresponding inverse processes we make use of the approximate detailed-balance treatment of \cite{OConnor2015a}. Below densities of $10^8\,$g\,cm$^{-3}$, we turn off all pair-process related source terms in the neutrino moment equations, but, in order to still be able to follow energy- and momentum-deposition in the low-density polar funnels, we apply the corresponding source terms for pair annihilation in the hydro equations (see \cite{Just2016} for more details on their computation).

For each simulation, $10^4$ equal-mass, passive tracer particles are initially placed in the disk, following the density distribution. The particles that exceed $r=10^9$\,cm are considered part of the outflow and set aside for post-processing (cf. \ref{sec:nucl-kilon-post}), with typically $\sim 2000$ outflow trajectories per model. All outflow particles remain within the computational domain for the duration of the simulations ($t=10$\,s).

\subsubsection{FLASH}

FLASH is a multi-physics simulation framework for astrophysical fluid dynamics \cite{fryxell00,dubey2009}.
To simulate long-term disk outflows in viscous hydrodynamics, we use the
dimensionally-split Piecewise Parabolic Method (PPM, \cite{colella84}) solver, which is based on the PROMETHEUS code 
as implemented in FLASH version 3.2. The public version has been modified to allow
for a non-uniform grid \cite{F12}, inclusion of a viscous stress in axisymmetry \cite{FM12},
and the pseudo-Newtonian potential of Artemova \cite{artemova1996} for gravity as reported in \cite{FKMQ14}. 

The neutrino implementation consists of a leakage scheme for cooling, with a local prescription
to compute the optical depth using the pressure scale height \cite{MF14,lippuner_2017}. 
Absorption is included using a lightbulb-type scheme that accounts for the annular
geometry of the accretion disk \cite{FM13}. Three neutrino species are included $(\nu_e,\bar{\nu}_e,\nu_x)$,
with the latter representing all 4 heavy lepton species. Charged-current weak interaction for emission
and absorption reactions of $\{\nu_e,\bar{\nu}_e\}$ with nucleons are included using the rates of \cite{Bruenn1985}.
Additionally, neutrino emission from  $e^+e^-$ pair annihilation and plasmon decay is included,
as well as opacity contributions from charged-current and neutral-scattering contributions
following \cite{ruffert_1996}, as reported in \cite{fernandez_2022_FFI}.

The neutrino physics in the FLASH-based setup have been gradually improved over time, resulting in quantitative
variations in the $Y_e$ distribution of the outflow that nevertheless do not
affect the results qualitatively. In \cite{FM13}, neutrino emission contained an exponential suppression of
emission with optical depth. In \cite{MF14}, this was replaced by a leakage scheme that computed
the production and diffusion times to modulate emission. In \cite{lippuner_2017}, separate mean energies
were computed for electron neutrino and antineutrino absorption from the disk (previously only
a single average energy was used for both species). Finally, since \cite{fernandez_2022_FFI},
the contributions of heavy lepton neutrinos are added in the leakage scheme, 
and the luminosity used for absorption in the next time step, computed from emission in the previous
time step, is corrected by the power absorbed in the previous time step (this leads to the 
correct hierarchy of number luminosities of electron-type neutrinos and antineutrinos).

By default, the equation of state is that of \cite{timmes2000}, with the abundances of neutrons, protons,
and alpha particles in nuclear statistical equilibrium (NSE), accounting for the nuclear binding energy
of alpha particles. An additional set of models is evolved with the same equation of state, but now
additionally including a heavy nucleus (${}^{54}$Mn) in nuclear statistical equilibrium, to capture
the additional nuclear energy release and match the EOS used by ALCAR (see Appendix~\S\ref{s:nse_appendix}).

The computational domain spans the radial range $[10^6,10^{11}]$\,cm and the full range of polar
angles, using a logarithmic grid in radius with 640 cells, and a polar grid equispaced in $\cos\theta$
with 112 cells ($\Delta r/r\simeq \Delta\theta\simeq 0.02$ at the equator). The boundary conditions
are outflow in radius and reflecting in polar angle.

FLASH models evolve tracer particles for post-processing in same way as the ALCAR models; see \S\ref{s:alcar}. 

\subsection{Nucleosynthesis and kilonova post-processing}\label{sec:nucl-kilon-post}

We employ a nuclear reaction network that includes 7362 nuclei from nucleons to $^{313}$Ds. We include $\alpha$-decay, $\beta$-decay, charged particle reactions, neutron captures and their inverse process, photo-disintegration, as well as spontaneous, neutron-induced, and $\beta$-delayed fission. It corresponds to the set of nuclear reactions labelled `FRDM' in ref.~\cite{Mendoza-Temis.Wu.ea:2015}.
We also consider weak interactions including the electron/positron captures and (anti-)neutrino absorption on nucleons.

For all trajectories, the nucleosynthesis calculation is started from the last time when the temperature reaches 10~GK. For each tracer the early evolution history of thermal quantities and weak interaction rates in the trajectory is obtained based on the simulation data. When the disk simulation ends at $t_f=10$~s, the tracer reaches a radius $r_f$.
After the end of simulation we take the assumption of homologous expansion such that the density is extrapolated as $\rho(t)=\rho(t_f) [1+v_f(t-t_f)/r_f]^{-3}$ with the asymptotic velocity $v_f$ at $t_f$. The temperature is evolved consistently, taking into account viscous and nuclear heating, and including the energy exchange associated with emission
and absorption of neutrinos.

Using the masses and final (i.e. at $t=t_f$) velocities of all trajectories, as well as the nuclear heating rates, mass fractions of lanthanides and actinides, and average mass numbers along each trajectory, we estimate the kilonova signal using the approach detailed in \cite{Just2022a}. The effective heating rates powering the kilonova are computed from the total heating rates using the approximate thermalization efficiencies of Ref.~\cite{Barnes2016a}. In addition to heating from $\beta^-$- and $\alpha$-decays as well as fission, which is treated following the standard treatment of~\cite{Barnes2016a} (as done in Ref.~\cite{Just2022a}), we find also a small contribution of $e^-$-capture and $\beta^+$-decays (see, e.g., Ref.~\cite{Wu2019c}) that are dominated by the decay of $^{56}$Ni, for which $80\,\%$ ($20\,\%$) of the energy goes into $\gamma$-rays (neutrinos). In contrast to the multi-dimensional kilonova analysis of \cite{Just2022a}, we assume spherical symmetry, as we are only interested in the most basic properties of the kilonova signal. To this end, we do not apply kernel-based interpolation techniques to map the trajectory properties to the velocity grid, but instead use simple 0th-order binning as follows: We discretize the velocity range between $v/c=0$ and $0.5$ using 50 bins and, for each velocity bin ranging from $v$ to $v+\Delta v$, obtain its mass $\Delta m$ by summation of all trajectories falling in this velocity range. The heating rates, lanthanide mass fractions, and average mass numbers (needed for the calculation of the gas energy density) for this bin are computed as mass-weighted averages over the same trajectories. The approximate radiative transfer equations are then solved on a finer grid (ranging from $v/c=0$ to 0.6 with 300 uniform zones) using linear interpolation to map from the coarser grid.

\begin{table*}[htb]
\caption{Models evolved and summary of results. The first three columns from the left show
model name (A: ALCAR, F: FLASH with no heavy nucleus in the EOS, Fh: FLASH with ${}^{54}$Mn), 
Shakura-Sunyaev viscosity parameter, and whether the standard neutrino species and emission processes
for each code are used (No = reduced to $\{\nu_e,\bar{\nu}_e\}$, with charged-current and neutrino-nucleon scattering
processes only). The following
two columns show total mass ejected at $r=10^9$\,cm as measured from the grid, and from
tracer particles, respectively, as well as average velocity from tracer particles at the same radius.
The last four columns show average quantities at the last position/time where the temperature is
$5$\,GK in each trajectory: electron fraction, entropy, radius, and expansion time (the latter
computed as average radius over average radial velocity at $5$\,GK).
\label{t:models}}
\begin{ruledtabular}
\begin{tabular}{lccccccccc}
Model & $\alpha$ & all $\nu$? & $M_{\rm ej,grid}$ & $M_{\rm ej,part}$ & $\langle v_r\rangle_{\rm r9}$ 
      & $\langle Y_e \rangle_{\rm 5GK}$ & $\langle S\rangle_{\rm 5GK} $ & $\langle r\rangle_{\rm 5GK}$ & $\langle t_{\rm exp}\rangle_{\rm 5GK}$ \\
      &          &           &($10^{-2}\,M_{\odot}$) & ($10^{-2}\,M_{\odot}$) & ($10^{-2}$\,c) 
      &                                 & (k$_{\rm B}$/b)               & ($10^7$\,cm)                 & (ms)                   \\
      
\noalign{\smallskip}
\hline
A-full-a6   & 0.06 & Yes & 1.87 & 1.81 & 4.19 & 0.256 & 19.0 & 6.33 & 75  \\
F-full-a6   &      &     & 2.06 & 2.12 & 3.73 & 0.283 & 19.4 & 6.96 & 79  \\
Fh-full-a6  &      &     & 2.26  & 2.30 & 4.27 & 0.279 & 17.4 & 7.31 & 77 \\
\noalign{\smallskip}                    
A-full-a3   & 0.03 & Yes & 1.54 & 1.50 & 3.56 & 0.277 & 20.0 & 5.56 & 98 \\
F-full-a3   &      &     & 1.67 & 1.67 & 2.87 & 0.298 & 20.0 & 6.57 & 102 \\
Fh-full-a3  &      &     & 1.91 & 1.88 & 3.27 & 0.289 & 18.6 & 6.78 & 101 \\
\noalign{\smallskip}                    
A-red-a3    & 0.03 & No  & 1.62 & 1.57 & 3.39 & 0.283 & 19.8 & 5.55 & 94  \\
F-red-a3    &      &     & 1.82 & 1.81 & 2.70 & 0.301 & 19.6 & 6.76 & 97  \\
Fh-red-a3   &      &     & 1.98 & 1.96 & 3.17 & 0.291 & 18.4 & 6.75 & 108 
\end{tabular}
\end{ruledtabular}
\end{table*}

\subsection{Model parameters}\label{sec:model-parameters}

The baseline configuration mirrors the parameters of model
\emph{m1} of \cite{Just2022_FFI},
which has a black hole of $3\,M_\odot$ and spin parameter of $0.8$. The
initial condition for the disk is an equilibrium torus with mass $0.1M_\odot$, 
constant initial $Y_e = 0.1$, constant entropy of $8$\,k$_{\rm B}$ per baryon, 
constant specific angular momentum, and 
radius of initial density maximum at $r=40$\,km. This equilibrium initial condition
is chosen to compare with previously published models and to facilitate analysis. A disk
formed self-consistently in a dynamical merger is expected to have a distribution of electron fractions,
entropies, and angular momenta, and the resulting outflow may differ from that obtained when
using an equilibrium torus (e.g., \cite{FQSKR-15,most_2021_disk,fujibayashi_2022,Just2023a}).
The kinematic viscosity coefficient follows
the functional form of \cite{shakura1973}, namely:
\begin{equation}
\nu = \alpha \frac{c_i^2}{\Omega_{\rm K}},
\end{equation}
where $\alpha$ is a constant, $c_i = \sqrt{p/\rho}$ is \emph{the isothermal sound speed},
with $p$ the gas pressure and $\rho$ the mass density, and $\Omega_{\rm K}$ is the equatorial 
Keplerian angular velocity of the pseudo-Newtonian potential. The default model has $\alpha=0.06$,
and we also consider alternative models with $\alpha=0.03$. Only the $r\phi$ and $\theta\phi$ components
of the viscous stress tensor are considered, in order to mimic conversion of shear kinetic energy
into thermal energy by turbulent angular momentum transport driven by the magnetorotational instability 
\citep{stone1999}.

Our naming convention prepends the letter ``A'' to models run with ALCAR, and ``F'' to models
run with FLASH. Models named \emph{full} use the entire production settings of each code as
described in \S\ref{s:codes},
with suffixes \{a3,a6\} when using $\alpha=\{0.03,0.06\}$, respectively. 

In addition, we evolve a set of models with reduced neutrino physics: no heavy lepton neutrinos, and
only charged-current neutrino/antineutrino emission and absorption, and neutral-current scattering on
nucleons. These models are denoted with ``red" (for reduced).

Also, a version of all models is repeated in FLASH, but now including a representative
heavy nucleus (${}^{54}$Mn, see Appendix~\ref{s:nse_appendix})
in the EOS, to match that used in ALCAR. These models start with ``Fh" (for heavy nucleus).

Finally, model F-full-a3 is repeated using 10 times more tracer particles than the default value, 
to test convergence of particle-based analyses (we name it F-full-a3-N10).

\section{Code comparison results \label{s:results}}

\subsection{Dynamics}\label{sec:dynamics}

\subsubsection{Accretion}\label{sec:accretion}

\begin{figure*}
\includegraphics*[width=\textwidth]{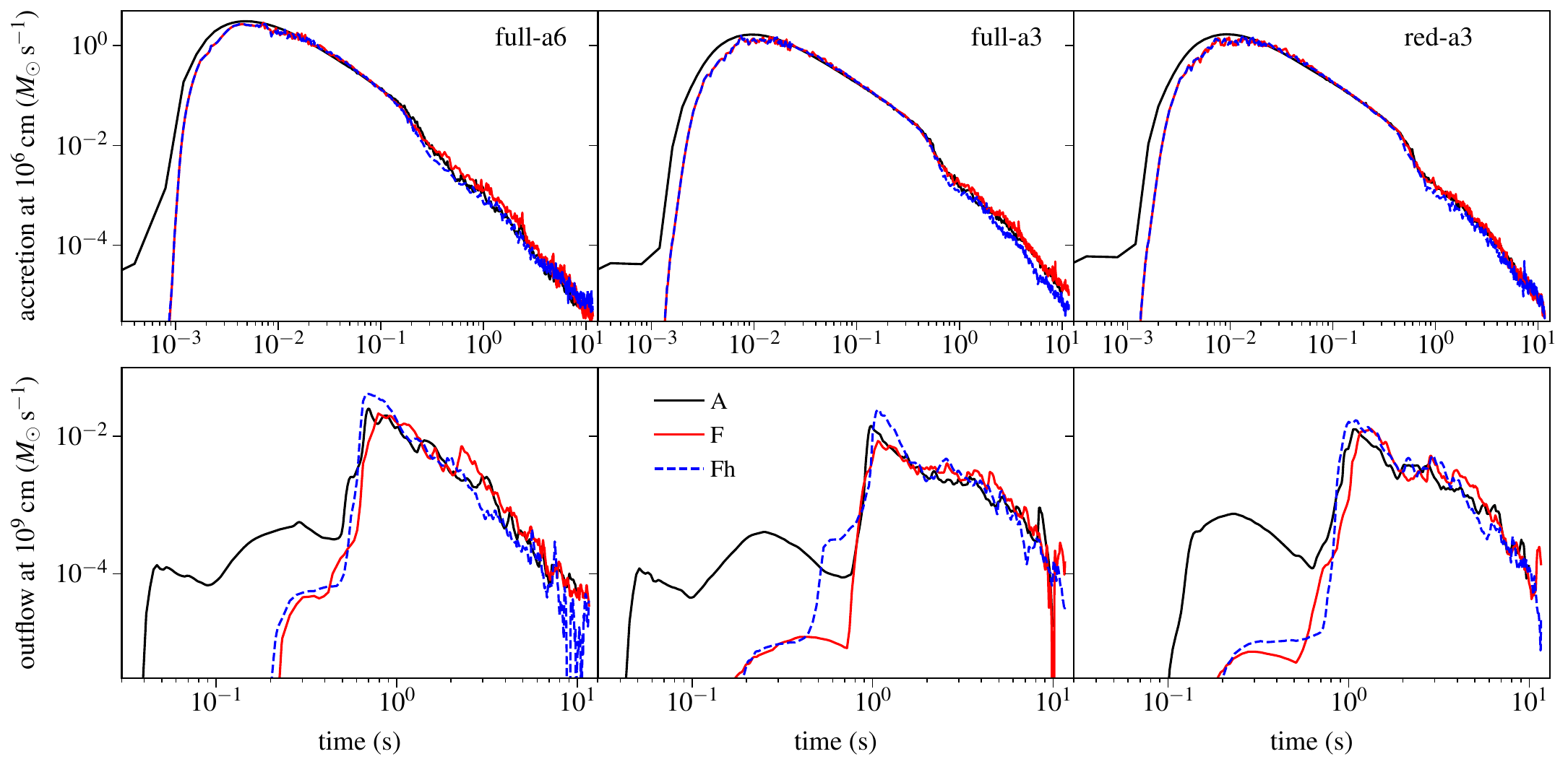}
\caption{Mass accretion rate at $r=10$\,km (top) and total mass outflow rate at $r=10^9$\,cm (bottom), 
for various models, as labeled for each column (see Table~\ref{t:models} for details). The offset
in the accretion rate at early times between ALCAR and FLASH models is due to the initial condition
and density floor values, which are not identical (despite the initial disk 
having the same physical parameters, 
see \S\ref{sec:model-parameters}). The bump in the ALCAR model outflow at $t\sim 200-400$\,ms
corresponds to the neutrino-driven wind, which we attribute to stronger neutrino absorption
in the M1 scheme relative to the lightbulb prescription used in FLASH.}
\label{f:mdot_mout_time}
\end{figure*}

The evolution of the disk during the first $\sim 100$ orbits at the radius
of initial density maximum ($\sim 300$\,ms) 
is mostly laminar,
and set by the interplay between viscous angular momentum transport and
neutrino cooling. Figure~\ref{f:mdot_mout_time} shows the mass accretion 
rate at $r=10$\,km, slightly inside the radius of the innermost stable circular orbit 
(ISCO, $\sim 13$\,km) for all
models evolved. Aside from a small offset in time, the evolution of the
accretion rate is nearly identical for ALCAR and FLASH models during
this initial phase. 

\begin{figure*}
\includegraphics*[width=\textwidth]{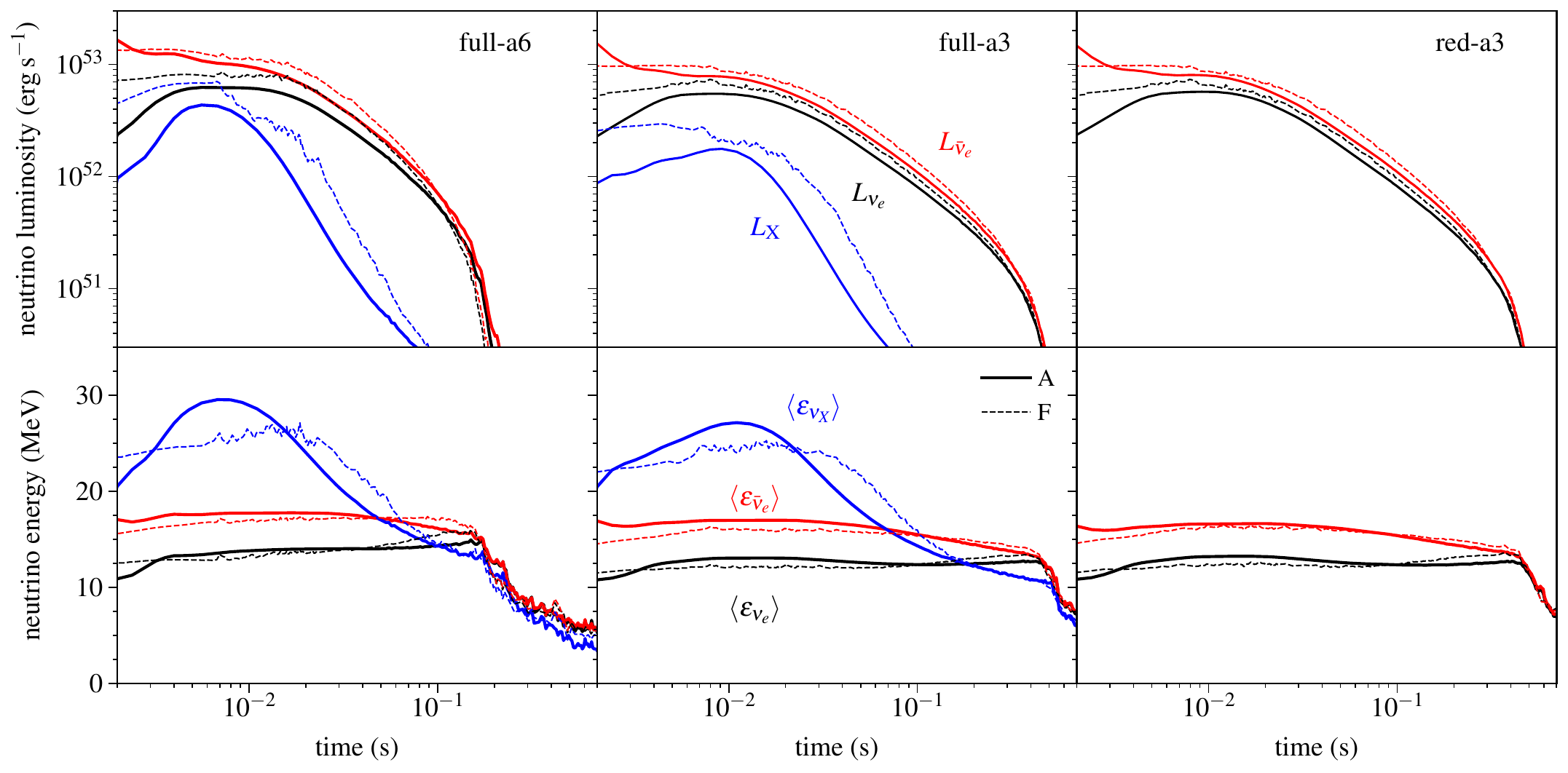}
\caption{Neutrino luminosities (top) and mean neutrino energies (bottom), for various models, as labeled for each column.
Quantities for A-models (ALCAR, solid lines, M1 neutrino scheme) are measured at $r=500$\,km, with
luminosities computed from the radiative flux evolved with the M1 scheme.
For 
F-models (FLASH, dashed lines, leakage+lightbulb neutrino scheme) luminosities are computed
as the sum of the (leakege-corrected) emissivities over the entire domain, corrected for absorption, hence the higher level of variability relative to A-models. The heavy lepton
luminosity includes the contribution from all $4$ species. F-models do not include the nuclear binding energy of a heavy nucleus.
}
\label{f:lum_ener_time}
\end{figure*}

Around $\sim 100$ orbits, with the exact value determined by the strength
of viscosity, neutrino cooling decreases sharply and the disk becomes
radiatively-inefficient. The timing of this transition at
 $\sim 200$\,ms and $\sim 450$\,ms
for $\alpha = 0.06$ and $0.03$, respectively (Figure~\ref{f:lum_ener_time}), shows excellent 
agreement between ALCAR and FLASH models. Combined with the
similarity of the inner accretion rate evolution, this agreement shows that
the viscous angular momentum transport is fully compatible
between the two implementations.

After the transition to the radiatively-inefficient (advective) phase, 
the disk becomes highly turbulent and the mass accretion rate at the ISCO 
becomes more stochastic. Figure~\ref{f:mdot_mout_time} shows
that the amplitude of fluctuations and overall evolution in accretion rate remains consistent
between ALCAR and FLASH models. At this stage, a small offset
becomes apparent between FLASH models that differ in the inclusion
of the nuclear binding energy of a representative heavy nucleus (models
F and Fh in Figure~\ref{f:mdot_mout_time}), with the models having a larger nuclear
binding energy release showing a larger drop in accretion rate (an effect first reported in \cite{lee_2009}).

\subsubsection{Outflow kinematics and nuclear energy release}\label{sec:outfl-kinem-nucl}

While the bulk of mass ejection in viscous hydrodynamic evolution 
takes place once the disk becomes radiatively inefficient,
earlier outflows do occur.
Figure~\ref{f:mdot_mout_time} shows the total mass outflow rate (bound and unbound) 
at $10^9$\,cm for all models. The most notable difference between ALCAR and
FLASH models is the early bump at $200-400$\,ms, which corresponds to mass ejection
driven by neutrino energy deposition (the ``neutrino-driven wind")\footnote{Note that there is
a finite travel time for outflow material to reach the extraction radius from the region where
it is launched: $10^9$\,cm$/0.1$\,c $\sim 0.3$\,s.}. 
This early outflow component is significantly larger in the ALCAR models,
which implement multi-group M1 neutrino transport. Unsurprisingly, the accuracy of
neutrino-driven mass ejection is dependent on the quality of the neutrino transport approximation.

Mass ejection following the transition to radiative inefficiency peaks at a time 
$\sim 1$\,s at the extraction radius located at $10^9$\,cm (Figure~\ref{f:mdot_mout_time}).
The rise, peak, and subsequent evolution of this component, which makes up the
majority of the disk wind, is similar yet quantitatively different for ALCAR and 
FLASH models. 

While identical evolution is not expected given the large stochatic 
fluctuations, a systematic difference is observed: both ALCAR
and FLASH models with a heavy nucleus (Fh) rise earlier to peak, reach a higher
peak, and decrease faster thereafter relative to the FLASH models (F) that only include
the nuclear binding energy of alpha particles. Table~\ref{t:models} shows that this
difference translates into a $\sim 10\%$ boost in ejected mass and 
a $10-20\%$ boost in average outflow velocity when comparing models F and Fh, which
only differ in the inclusion of the nuclear binding energy of ${}^{54}$Mn. Compared
to ALCAR models, F models have all lower average velocity but eject more mass. 

\begin{figure*}
\includegraphics*[width=0.49\textwidth]{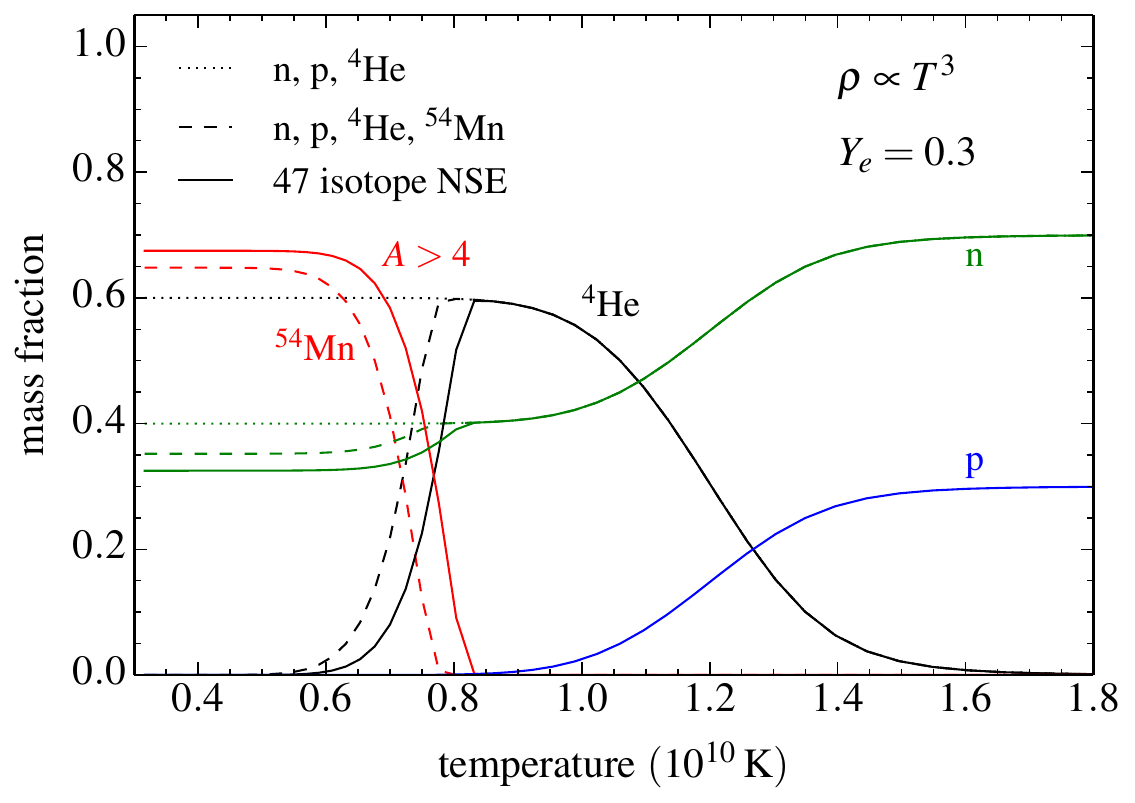}
\includegraphics*[width=0.49\textwidth]{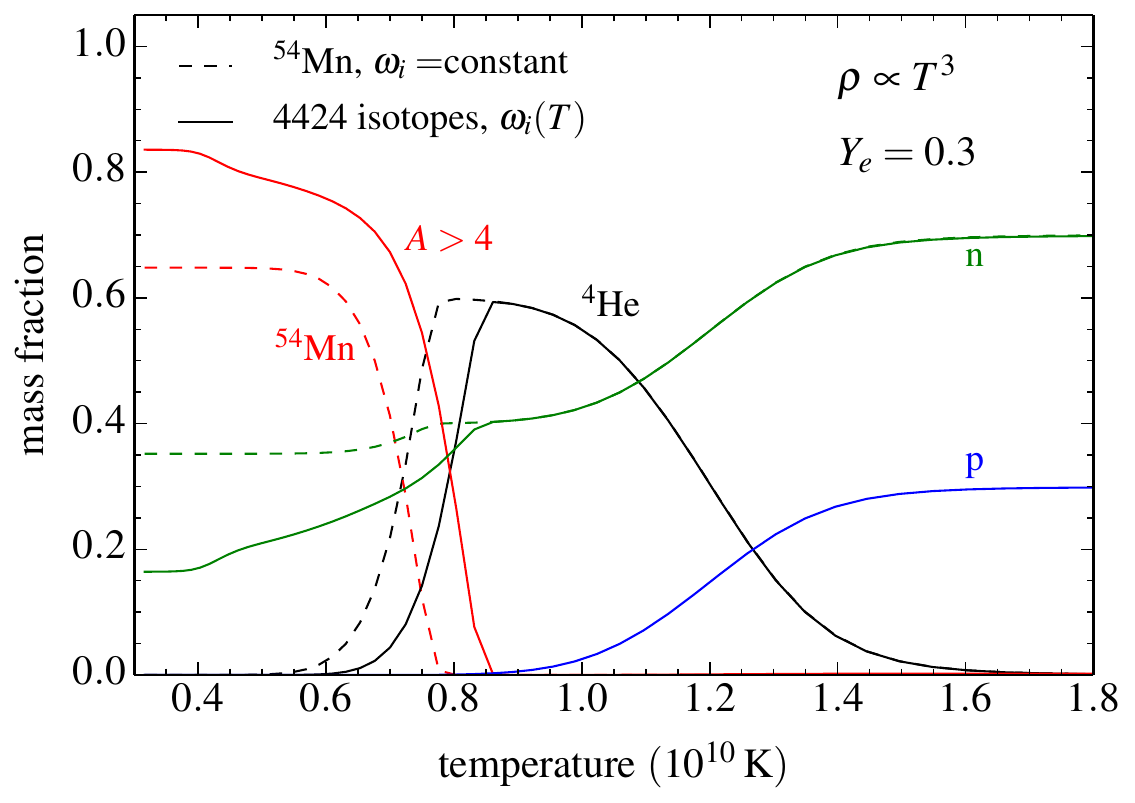}
\caption{\emph{Left:} Abundances in NSE as a function of temperature, for three different mixtures:
neutrons protons and alpha particles (dotted lines), adding ${}^{54}$Mn as the
only heavy nucleus (dashed), and the 47 isotope NSE mixture of \cite{seitenzahl_2008} (solid lines). 
The NSE equations are solved with constant partition functions, as in Appendix~\ref{s:nse_appendix}.
The electron fraction and thermodynamic path are chosen to be representative of
the disk outflow ($\rho\propto T^3$, with $T=3\times 10^8$\,K at $\rho=2000$\,g\,cm$^{-3}$). F models use an EOS
without a heavy nucleus (dotted lines), while Fh and A models use ${}^{54}$Mn as a representative
heavy nucleus. The red solid curve contains the mass fractions of all nuclei heavier than ${}^{4}$He, the
asymptotic value at low temperature is 0.675. For the dashed line, this value is 0.648.
\emph{Right:} Same as in the left panel, now comparing the same mixture with ${}^{54}$Mn and constant
partition functions as in the left panel (dashed lines), with a mixture of 4452 isotopes that uses temperature-dependent partition functions in the NSE equations (solid lines). The asymptotic value of the solid red curve at low temperature
is 0.836.}
\label{fig:nse}
\end{figure*}

To illustrate the magnitude of the nuclear energy release in the different EOS mixtures,
we plot in Figure~\ref{fig:nse} the NSE abundances for a representative thermodynamic 
path of the outflow ($Y_e=0.3$, $\rho\propto T^3$). At low temperature, the difference
in nuclear binding energy released per nucleon between the Fh (${}^{54}$Mn) and F (${}^{4}$He only) 
models is
\begin{equation}
(0.648\times 8.74-0.6\times 7.07)\ \text{MeV}\simeq 1.4\ \text{MeV}. 
\end{equation}
If fully converted to kinetic energy, this would correspond to a specific kinetic energy of
$2.9\times 10^{-3}$\,c$^2$. For an initial velocity of $0.035$\,c, the additional kinetic energy would 
boost the ejecta to $0.065$\,c. The difference in expansion velocity at $10^9$\,cm between F and Fh models 
($\sim 0.005$\,c) is much smaller than this value, however, indicating that most of this additional 
nuclear energy released remains as thermal energy at least up to this radius.

\begin{figure*}
\includegraphics*[width=\textwidth]{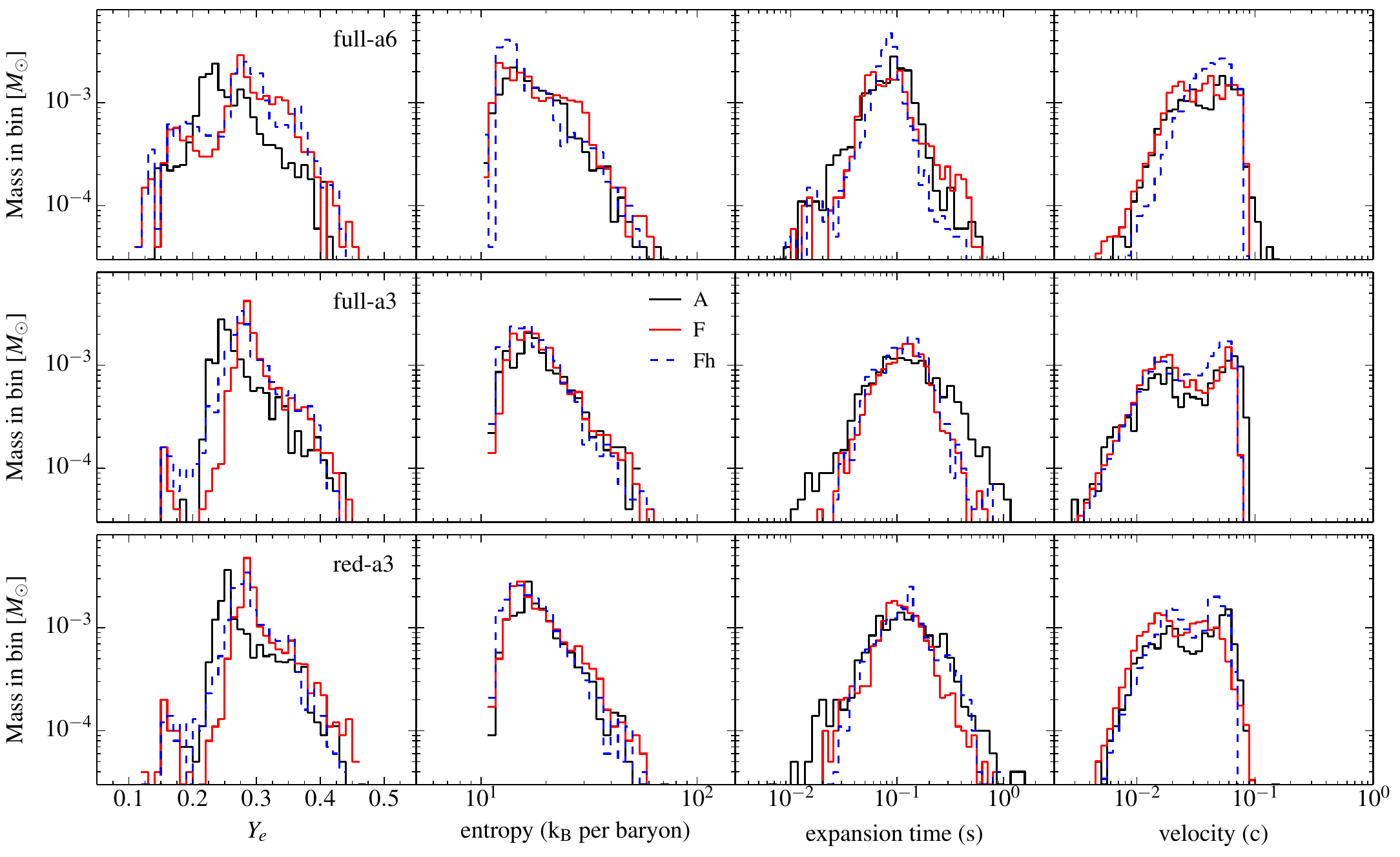}
\caption{Mass histograms as a function of various quantities, for the 3 different model configurations we explore, 
as labeled for each row. Quantities are computed from tracer particles that are ejected past $r=10^9$\,cm by the end of the
simulation at $t=10$\,s. The electron fraction $Y_e$, entropy, and expansion time $t_{\rm exp}=r/v_r$ are computed
at the last time a particle reached $T=5$\,GK. For the latter, we only include particles with positive velocity.
The velocity histogram is computed at $r=10^9$\,cm. }
\label{f:histograms}
\end{figure*}

The outflow velocity distribution is shown in the rightmost column of Figure~\ref{f:histograms}.
In all cases, the velocity distribution has the same qualitative form: a double-peaked structure, 
a sharp cutoff at $\sim 0.1$\,c, and an extended tail to lower velocities. The distribution
shows excellent agreement between all models that use $\alpha=0.03$ (full-a3 and red-a3), with quantitative
differences related to the amount of mass ejected. A more noticeable
difference appears in the full-a6 set, for which model Fh shows low-velocity tail that is
shifted to higher velocities. This is consistent with the larger average velocity shown
in Table~\ref{t:models} which is due to less material moving at low speeds.
Overall, ALCAR and FLASH models that include the nuclear binding energy
contribution from ${}^{54}$Mn (Fh) have average velocities that differ by less than
$10\%$, thus accounting for most of the difference between A and F models.

Figure~\ref{fig:nse} also shows abundances as a function of temperature
for two NSE mixtures with more nuclei. Using the 47 isotope mixture
of \cite{seitenzahl_2008} and employing constant nuclear partition
functions $\omega_i$, results in a marginally higher abundance of heavy nuclei
(everything other than ${}^{4}$He, n, p) by $4\%$ relative to using only ${}^{54}$Mn, 
with an increase in the nuclear
energy release of the same magnitude. Significant differences
require a much larger number of isotopes: the right panel of Figure~\ref{fig:nse}
shows abundances obtained with a mixture of 4452 isotopes and using
temperature-dependent partition functions $\omega_i(T)$.
While the effect of the temperature-dependent partition functions is to shift the 
transition between nuclear species to slightly higher temperature
relative to the case of constant partition functions, the larger number
of nuclei allows to reach a higher mass fraction and consequently larger
nuclear energy release. The $20\%$ increase in heavy nuclei mass fraction results in 
an extra $\sim 1.7$\,MeV per nucleon released. 
In terms of additional kinetic energy
when fully converted, and relative to using only ${}^{54}$Mn as a representative nucleus
and $0.035$\,c as a baseline velocity when only including alpha particles, 
this extra nuclear energy release would boost the  outflow from $0.065$\,c to $0.088$\,c.
This motivates future work toward improving
how nuclear physics and $r$-process heating is included in post-merger simulations.

\subsubsection{Neutrino quantities and equilibrium $Y_e$}\label{sec:neutr-quant-equil}

The neutrino luminosities and mean energies for A- and F-models are shown in Figure~\ref{f:lum_ener_time}. 
In ALCAR, the M1 luminosities are measured at $500$\,km, whereas in FLASH they
are computed instantaneously in the entire domain, correcting for the neutrinos absorbed, as in \cite{fernandez_2022_FFI}.
Despite the different transport methods, the global electron-type neutrino and antineutrino luminosities after $t\sim 3$\,ms
are consistent in both codes to within 10--20\%, regardless of the neutrino physics included (i.e., model full-a3 versus red-a3). 
A larger discrepancy of up to a factor $\sim 2$ is obtained
in the heavy-lepton luminosities (models full-a6 and full-a3). Nevertheless the time evolution is remarkably close in all species, 
owing to the agreement in angular momentum transport and global dynamics as discussed in \S\ref{sec:accretion}.

Figure~\ref{f:lum_ener_time} also shows the mean energies for all neutrino species evolved,
obtained as the global ratio of energy- to number luminosities for each species (as in \cite{ruffert_1996}).
For electron-type neutrinos and antineutrinos, the mean energies show close similarity as with the luminosities,
with no significant differences in the level of agreement between models that include all neutrino interactions
and those that reduce the neutrino emitting channels. Again, a larger discrepacy is observed in the mean
energies of heavy lepton neutrinos.

\begin{figure*}
\includegraphics*[width=\textwidth]{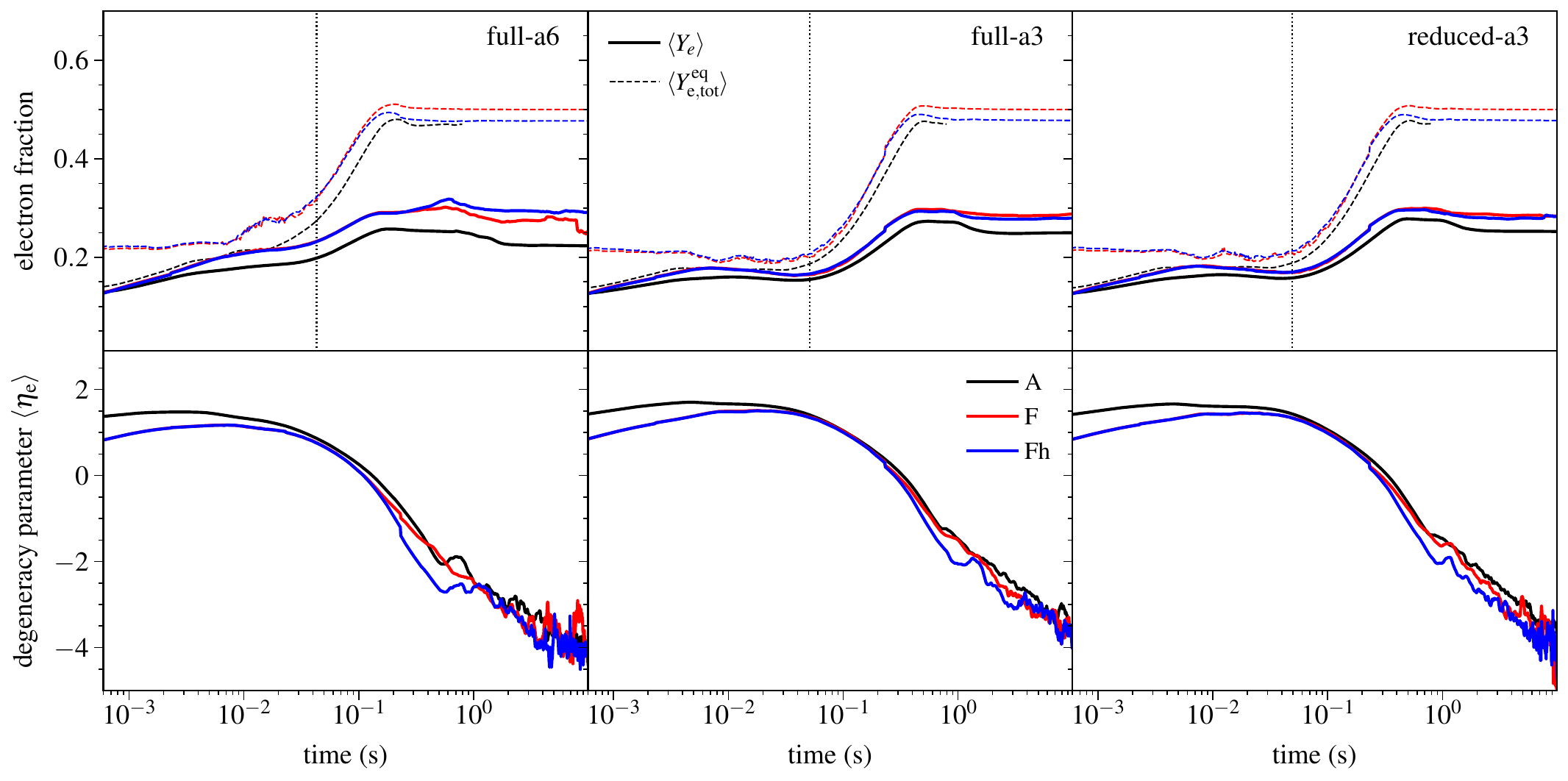}
\caption{\emph{Top:} mass-averaged electron fraction (solid lines) and mass-averaged equilibrium electron fraction
in the disk, considering both emission and absorption (dashed lines).
The average considers matter that satisfies $T>0.1$\,MeV, $\rho>10^{3}$\,g\,cm$^{-3}$, and $r < 10^9$\,cm. 
The vertical dotted line indicates the time at which $L_{\nu_{\rm e}}$ drops by a factor 3 from its maximum value in ALCAR models
(c.f. Figure~\ref{f:lum_ener_time}).
\emph{Bottom:} mass-averaged
electron degeneracy parameter, excluding rest mass. Note that, while the initial condition in ALCAR 
and FLASH models has the same physical parameters, it is constructed independently using slightly different
equations of state, which accounts for the initial offset in equilibrium $Y_e$ and degeneracy parameter.}
\label{fig:yeq_eta}
\end{figure*}

The electron fraction distribution of the outflow at $T=5$\,GK (Figure~\ref{f:histograms}) shows a systematic shift of its peak 
toward lower electron fractions by $\sim 0.02-0.03$ in ALCAR models relative to FLASH models, consistent with the
offset in average $Y_e$ shown in Table~\ref{t:models}. The entropy distribution peaks at lower values in FLASH models,
but shows otherwise a similar shape relative to ALCAR models, consistent with the agreement in mean
values (Table~\ref{t:models}). The expansion time also shows consistent distributions between FLASH and ALCAR models,
with larger deviations in Fh models relative to both F and A models.

We can analyze the offset in $Y_e$ by computing the equilibrium values toward which weak interactions are driving
the composition in the disk. These equilibrium values are obtained by balancing the rates of neutrino
and antineutrino emission/absorption (as in, e.g., Ref.~\cite{Just2022_Yeq}). We denote by $\langle Y_{\rm e,em}^{\rm eq}\rangle$
the mass-averaged equilibrium electron fraction obtained by balancing electron neutrino and antineutrino
emission rates, $\langle Y_{\rm e,abs}^{\rm eq}\rangle$ the corresponding equilibrium value obtained with absorption
rates only, and $\langle Y_{\rm e,tot}^{\rm eq}\rangle$ the equilibrium value obtained by balancing both emission and absorption rates.

\begin{figure}
\includegraphics*[width=\columnwidth]{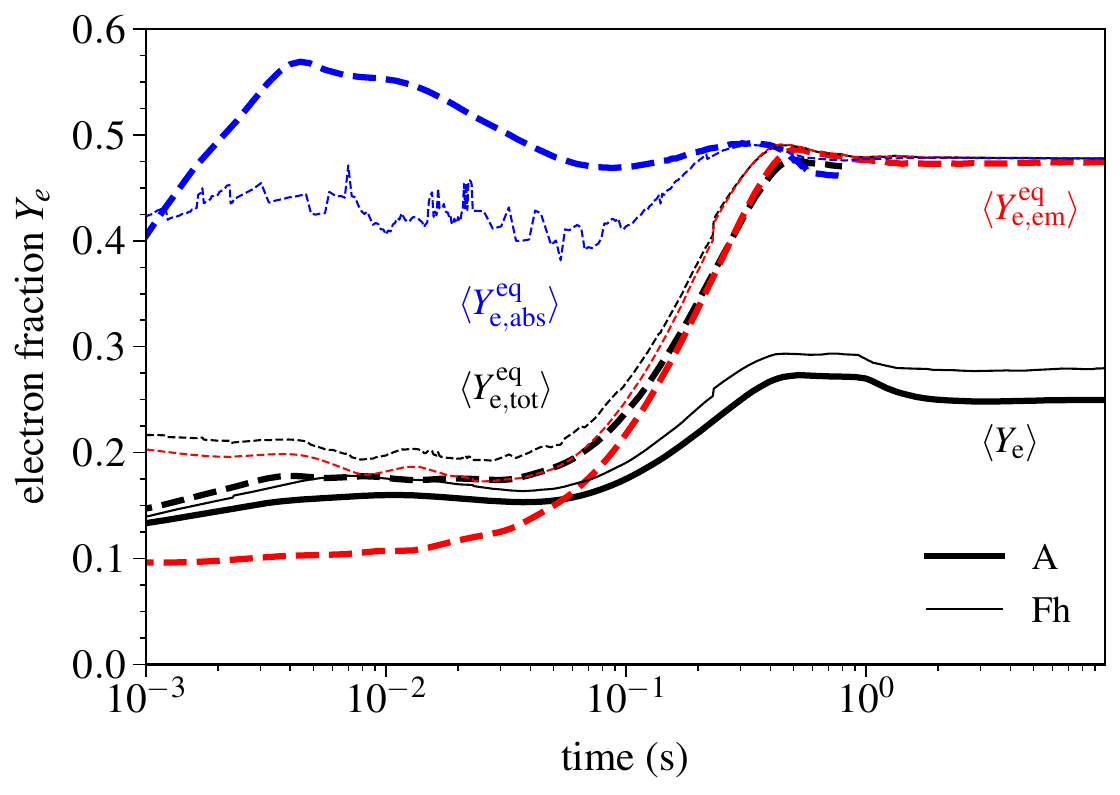}
\caption{Mass-averaged electron fraction (solid lines) 
and equilibrium electron fractions (dashed lines), as labeled,
for models A-full-a3 (thick lines) and Fh-full-a3 (thin lines). }
\label{fig:yeq_components}
\end{figure}

Figure~\ref{fig:yeq_eta} shows the evolution of the mass-averaged electron fraction $\langle Y_{\rm e}\rangle$ and the
total equilibrium value towards which weak interactions are driving it. At early times, these equilibrium 
values are moderately neutron rich ($\sim 0.2$), consistent with the mild electron degeneracy of the disk (also shown 
in Figure~\ref{fig:yeq_eta}). 
The offset at $t=0$ in equilibrium $Y_e$ and electron degeneracy parameter is 
due to the different
treatment of the ion entropy in each EOS. In FLASH, the entropy of ions is obtained by assuming that
they are a single species with average mass number $\bar{A}$ and adiabatic index $5/3$, 
ignoring the statistical weight $g$ (spin degree of
freedom), see equation~(110) in Ref. \cite{fryxell00}. 
Since both protons and neutrons have $\bar{A}=1$, they are treated as the same
species and thus, in addition to ignoring the difference between their masses, 
the ``entropy of mixing" (terms of the form $Y_i \ln (Y_i)$, with $Y_i$ the number fraction) is absent.
In ALCAR, the ion entropy is computed as the sum of individual entropies from
the different species considered, 
and including statistical weights ($g=2$ for nucleons and $g=1$ for nuclei). This results in a difference of
$\sim 1$\,k$_{\rm B}$ per baryon in total entropy between the two EOSs for a given $\{\rho, T, Y_e\}$ 
combination characteristic of the initial torus density maximum. Since both codes start
from equilibrium tori constructed with an entropy of $8$\,k$_{\rm B}$ per baryon, 
the thermodynamic conditions have a small initial offset.

As the disk density decreases and degeneracy drops, weak  equilibrium increases $Y_e$  toward 
$\sim 0.5$ because free nucleons recombine into $\alpha$ particles and heavy nuclei (characterized by an average mass number $A_{\rm h}$ and charge number $Z_{\rm h}$). 
Assuming full recombination and a fixed representative heavy nucleus, mass and charge conservation lead to the following relation for the asymptotic equilibrium electron fraction at low temperature: 
\begin{equation}
\label{eq:asymptotic_ye}
Y^\text{eq}_e = \frac{1}{2} - \left[\frac{1}{2}-\left(\frac{Z_{\rm h}}{A_{\rm h}}\right)\right]\,X_{\rm h} \, ,
\end{equation}
where Eq.~(\ref{eq:asymptotic_ye}) is also valid in the case where $X_{\rm h}, A_{\rm h}$, and $Z_{\rm h}$ denote the (average) properties of a distribution of heavy nuclei.

Consistent with these considerations, both ALCAR models and FLASH models that assume ${}^{54}$Mn as representative heavy nucleus have on average $X_{\rm Mn}\sim 0.6$ at late times, i.e. $Y^\text{eq}_e \simeq 0.48$, that corresponds to  
the asymptotic average $\langle Y_e^{\rm eq}\rangle$ at low temperatures. FLASH models without ${}^{54}$Mn, i.e. only $\alpha$ particles, have $Y^\text{eq}_e = 0.5$ (Figure~\ref{fig:yeq_eta}).

The mass-averaged electron fraction follows the shape of the equilibrium value without reaching it in both ALCAR and FLASH models, decoupling around the time at which neutrino luminosities drop significantly (cf. Figure~\ref{f:lum_ener_time}). An offset between ALCAR and FLASH models is apparent in both the average electron fraction and the equilibrium value, with FLASH models showing consistently higher values. The offset in the electron fraction distribution seen in Figure~\ref{f:histograms} thus most likely originates from the offset in the equilibrium electron fractions obtained with each code, given that all models start with the same initial electron fraction.

To analyze this offset further, we separate the equilibrium electron fraction into emission and absorption
components in Figure~\ref{fig:yeq_components} for the full-a3 models. At early times, the ALCAR model has a 
lower emission equilibrium and a higher absorption equilibrium than FLASH models, such that the net equilibrium value is lower. The difference in emission equilibrium can be attributed to the higher electron degeneracy in the ALCAR models (Fig.~\ref{fig:yeq_eta}), while the difference in absorption equilibrium is likely due to the different neutrino transport implementations.

\begin{figure}
\includegraphics*[width=\columnwidth]{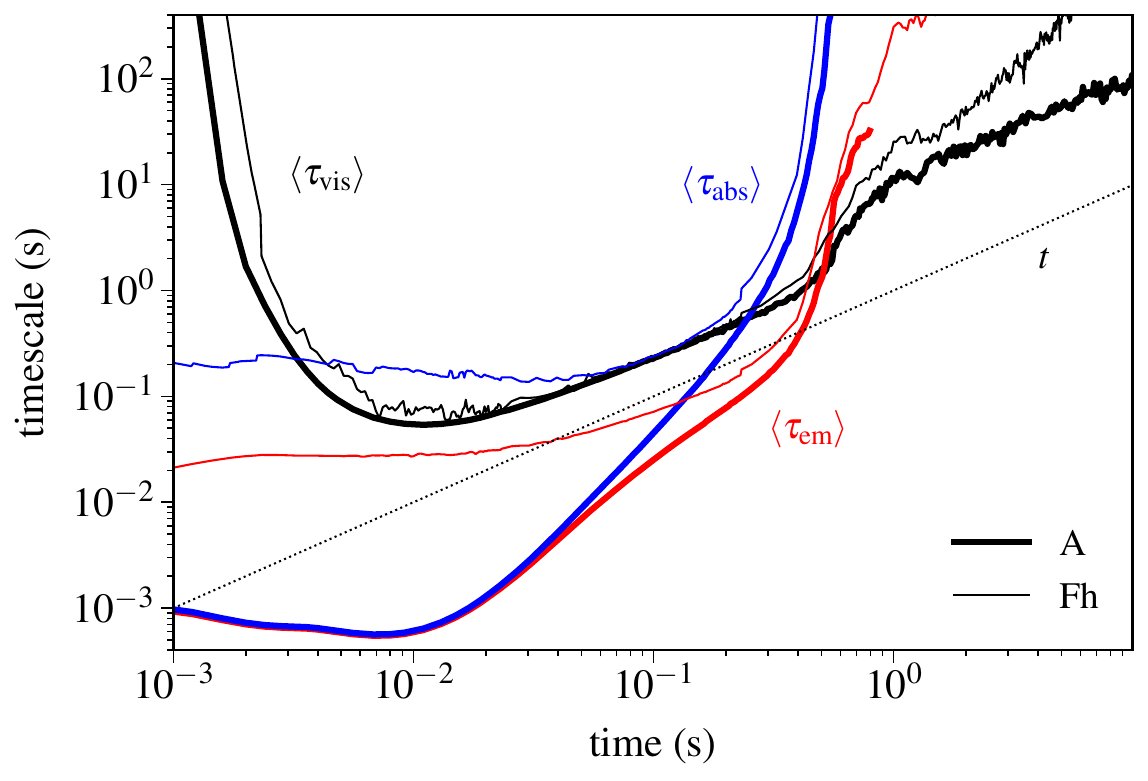}
\caption{Characteristic timescales governing the evolution of $Y_e$ in the disk for full-a3 models (ALCAR: thick
lines, FLASH: thin lines). Shown are the weak interaction
timescale due to neutrino emission $\langle \tau_{\rm em}\rangle$ and absorption $\langle \tau_{\rm abs}\rangle$,
and viscous (accretion) timescale $\langle \tau_{\rm vis}\rangle$, all computed as in Ref.~\cite{Just2022_Yeq}.
The dotted line denotes the current time $t$.}
\label{fig:torus_timescales}
\end{figure}

Note that the average $Y_e$ in FLASH models takes longer to approach the equilibrium electron fraction 
than in ALCAR models, because the effective weak interaction timescales are significantly longer 
due to the leakage and absorption
implementation. Figure~\ref{fig:torus_timescales} shows the average weak interaction timescales in the disk due to
neutrino emission $\langle \tau_{\rm em}\rangle$ and absorption $\langle \tau_{\rm abs}\rangle$, as well
as the accretion (viscous) timescale $\langle \tau_{\rm vis}\rangle$, for the full-a3 model, 
computed as in Ref.~\cite{Just2022_Yeq}.
In the ALCAR model, the emission and absorption timescales are shorter than the
accretion and current times $t$, hence the torus approaches $Y_e$ equilibrium quickly and remains
close to that state until freeze-out, as shown in Figure~\ref{fig:yeq_components}. The FLASH model, on the
other hand, is such that the shortest timescale, $\langle \tau_{\rm em}\rangle$, is initially
shorter than the accretion time but longer than the current time, hence the $Y_e$ of the 
torus remains out of equilibrium until $t\sim 10$\,ms.

As time elapses and the absorption contribution decreases, the net equilibrium $Y_e$ merges with
the emission equilibrium, and the offset in equilibrium value between the two codes decreases. Despite these
differences, the actual mass-averaged electron fraction between the two codes has a moderate offset of $\sim 0.02$
throughout the evolution.

\subsection{Nucleosynthesis}\label{sec:nucleosynthesis}

\begin{figure*}
\includegraphics*[width=\textwidth]{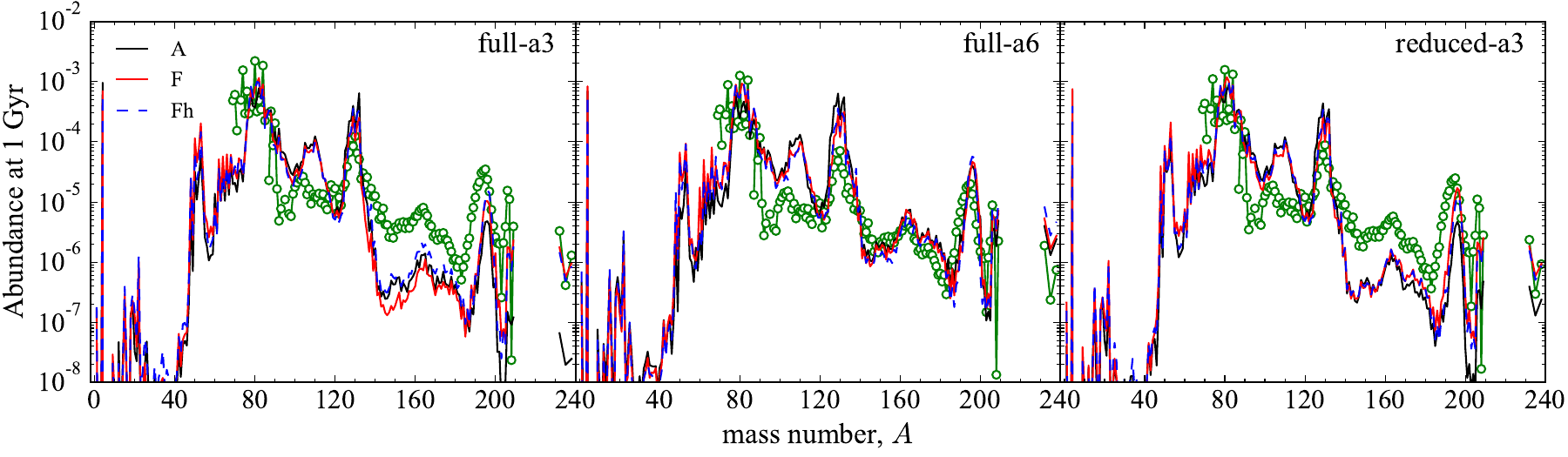}
\includegraphics*[width=\textwidth]{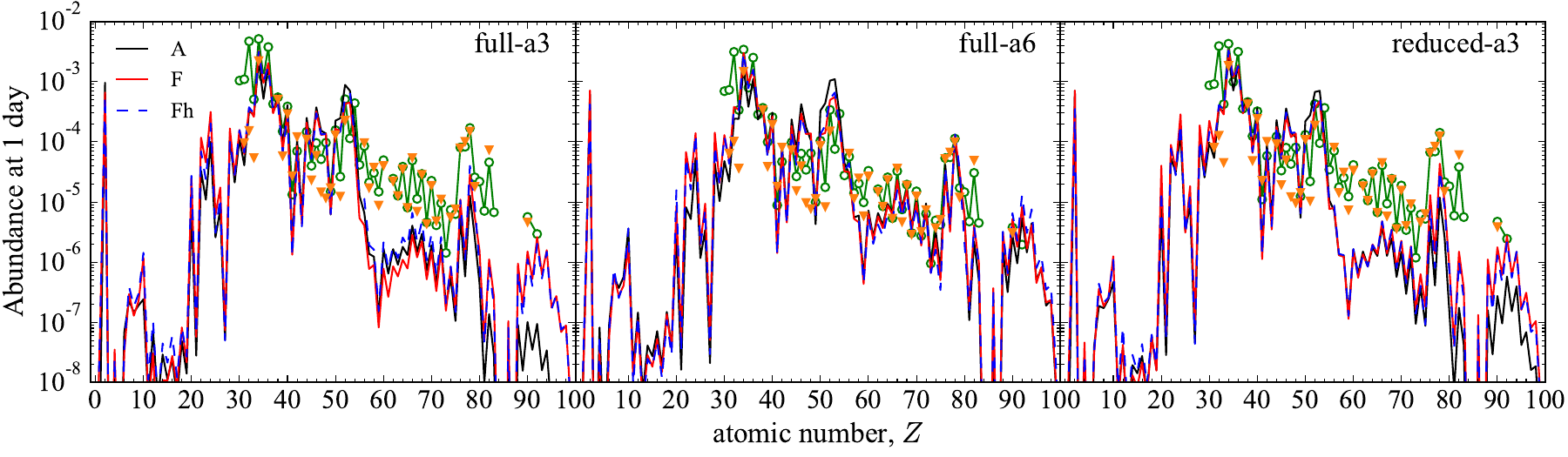}
\caption{Abundance yields as functions of mass number $A$ at 1 Gyr (top row) and atomic number $Z$ at 1 day (bottom row). Green circles show the solar $r$-process abundances \cite{Goriely1999} and are scaled to the yields of Sr in the ALCAR models. Orange triangles in the lower row show abundances observed for the metal-poor star HD-222925 \cite{Roederer2022} and are scaled to the solar Eu abundance.
}
\label{f:nusyn}
\end{figure*}

Figure~\ref{f:nusyn} compares the abundance yields from ALCAR and FLASH models as functions of mass number $A$ at 1\,Gyr and of atomic number $Z$ at 1\,day. The abundance patterns in models F and Fh are very similar to each other, except that lanthanides are enhanced by a factor of $\sim$2--4 in model Fh-full-a3 compared to model F-full-a3.
This suggests that the nucleosynthesis outcome is not very sensitive to the inclusion of the nuclear binding energy of $^{54}\rm Mn$ (Fh).

\begin{figure*}
\includegraphics*[width=\textwidth]{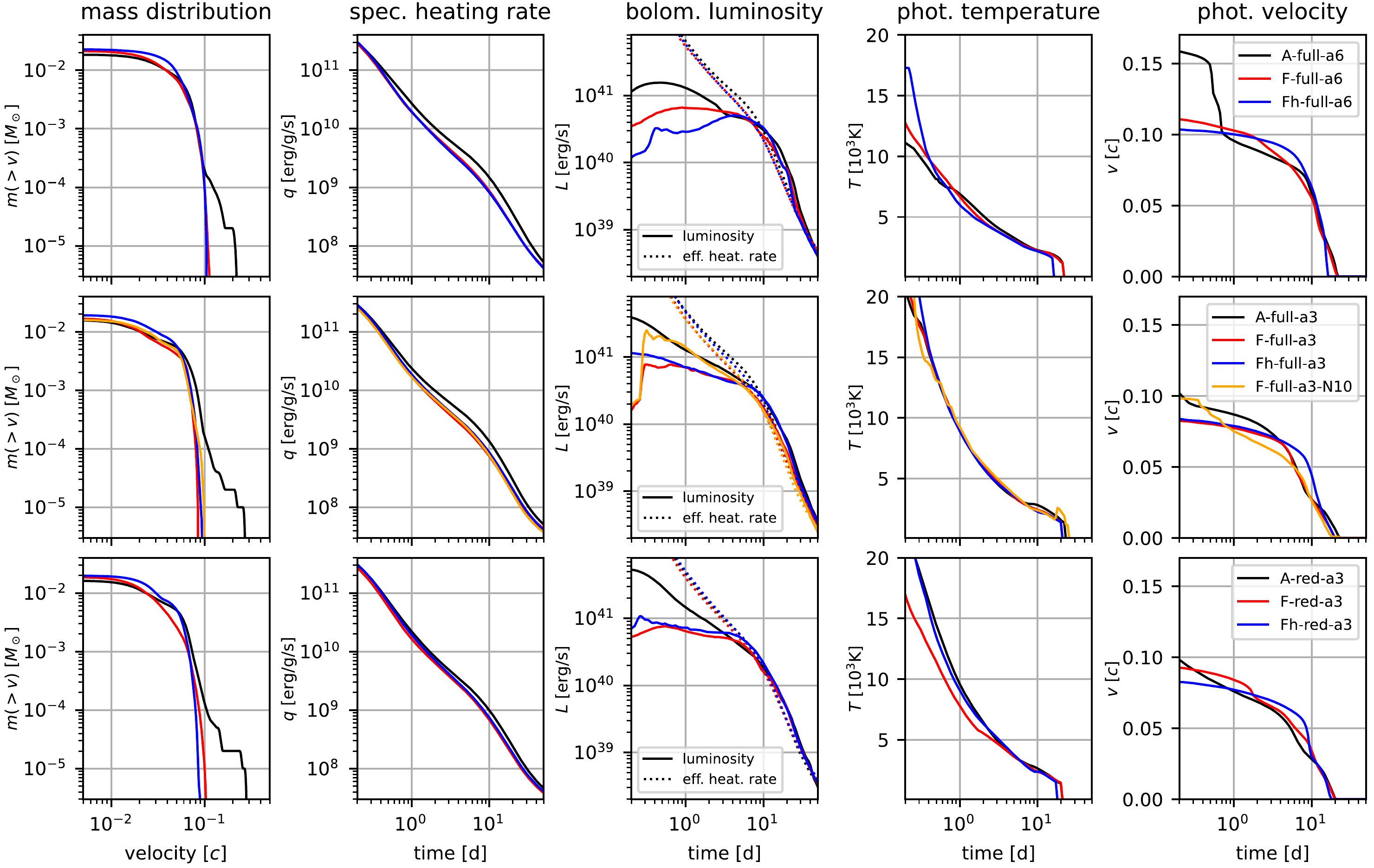}
\caption{Basic kilonova properties, namely, from left to right, the mass distribution (i.e. for given $v$ the mass of material faster than $v$), the specific heating rate (before thermalization and including neutrino contributions), bolometric luminosity (solid lines) and total effective heating rates (dotted lines), photospheric temperature, and photospheric velocities. The last two quantities are computed as in Eqs.~(28) and~(29) of \cite{Just2022a}. Each row shows results for the models listed in the right panel.}
\label{f:kilonova}
\end{figure*}

Overall, ALCAR and FLASH models agree very well. The abundance patterns near the first $r$-process peak are well reproduced compared to that in the metal-poor star HD-222925 \cite{Roederer2022}. Consistent with the offset for the peak electron fraction from $\sim 0.28$--0.29 in the FLASH models to $\sim 0.23$--0.24 in the ALCAR models (Figure~\ref{f:histograms}), the abundances of 2nd-peak $r$-process elements ($A\approx 130$) in the ALCAR models are higher than in the F and Fh models, namely by a factor of $\sim$1.5--2, because those elements are most efficiently produced from the ejecta with $Y_e=0.2$ to 0.23. The enhancement of $^{132}\rm Te$ at 1 day leads to higher specific heating rates (shown in the second column of Fig.~\ref{f:kilonova}) originating from the $\beta^-$ decay chain of $^{132}\rm Te$--$^{132}\rm I$--$^{132}\rm Xe$. A noteworthy difference is observed in the abundance of actinides, which is about a factor of 10--40 smaller in the ALCAR models, associated with the lack of very neutron-rich ejecta with $Y_e<0.2$ in these models. The small amount of ejecta with $Y_e\sim 0.1$ in FLASH models (Figure~\ref{f:histograms}) is apparently sufficient to make a significant difference in the yields of actinides, therefore potentially allowing these elements to be used as diagnostics of the ejecta electron fraction.

The model sets full-a3 and red-a3 underproduce nuclei with $A>140$ compared to the solar $r$-process pattern, while the model set full-a6 shows a more consistent abundance pattern even in 3rd-peak elements and actinides. As found in previous studies (e.g. \cite{FM13,Just2015a,Just2022_Yeq}), a higher viscosity leads to faster matter ejection and, therefore, earlier freeze-out of $Y_e$, increasing the fraction of matter ejected with $Y_e<0.2$ (i.e. neutron-rich enough to enable actinide production). The early freeze out results in more matter being ejected with $Y_e < 0.2$, leading to a more solar-like distribution of elements heavier than $A\approx 140$ \cite{Just2022_Yeq}.

\subsection{Kilonova signal}\label{sec:kilonova-signal}

The kilonova signal produced by the ejecta is compared for all models in Figure~\ref{f:kilonova}. As anticipated from the similarity of outflow properties and nucleosynthesis yields, the basic kilonova properties (bolometric luminosity, photospheric temperature and velocity) show an overall good level of agreement, especially considering that matter ejection from turbulent disks involves a non-negligible level of stochasticity.

The specific heating rates, shown in the second column from the left, differ only marginally when varying the physics input, while they are systematically shifted to higher values (about 20--40\,\% during the considered times) in the ALCAR models compared to the FLASH models. This difference is connected to the more pronounced 2nd $r$-process peak in the ALCAR models, as mentioned in the previous section. However, the impact of this difference on the brightness of the kilonova is partially compensated by the slightly smaller ejecta masses of the ALCAR models.

Since BH-disk ejecta are relatively slow compared to other ejecta components that can be produced in a NS merger, they become optically thin at rather late times, $t\sim 5\textit{--}10\,$d. These transition times when the ejecta start to become optically thin can be read off in the third column of Figure~\ref{f:kilonova} as the times when the total luminosities for the first time exceed the effective heating rates. Previous to these transition times, most of the ejecta are still optically thick and the emission is produced from just the outermost ejecta layers (see first and fifth column of Fig.~\ref{f:kilonova}). The higher luminosities seen at early times in the ALCAR models are likely to be connected to the more pronounced neutrino-driven mass ejection in the ALCAR models, which leads to an extended high-velocity tail.

However, another, purely numerical reason may also be poor sampling of the high-velocity edge of the ejecta with tracer particles. This is suggested by the comparison with model F-full-a3-N10, which evolves 10 times more tracer particles than model F-full-a3 and exhibits significantly brighter emission at early times. This comparison suggests that the adopted number of $1500-2000$ equal-mass tracer particles is high enough to describe the main part of the light curve, during which the photosphere travels through the ejecta, but is insufficient for accurately resolving the emission at earlier times. We note, however, that the reduced accuracy at early times may not be overly relevant for kilonova modeling of NS mergers, because the early light curve is likely to be dominated by other ejecta components.

\section{Summary and discussion \label{s:discussion}}

We have carried out a code comparison study of ALCAR and FLASH,
both of which have been used extensively over the past decade to study
the viscous hydrodynamic evolution of BH accretion disks formed
in neutron star mergers. For the comparison, we employ 
a representative system around a BH with identical initial conditions, and vary the viscosity
parameter as well as neutrino physics. Our main results are the following:
\newline

\noindent
1. -- We find excellent agreement in the quantities that
depend on angular momentum transport, i.e., the accretion rate history,
and timing of weak interaction freeze-out (Figures~\ref{f:mdot_mout_time} and \ref{f:lum_ener_time}).
A larger discrepancy is obtained in quantities that depend on the neutrino
transport approximation, such as the magnitude of the neutrino-driven wind (Figure~\ref{f:mdot_mout_time}) and
the electron fraction distribution of the ejected material (Figure~\ref{f:histograms}).
\newline

\noindent
2. -- Both codes show the same progression of the equilibrium electron fraction
from low to high values as the disk becomes less degenerate over time. Slightly
higher electron degeneracies, and therefore more neutron-rich equilibrium conditions,
are found in the ALCAR models, which accounts for the $\sim 10\%$ shift to lower average
$Y_e$ values in the ejecta of the ALCAR compared to the FLASH models 
(Figs.~\ref{fig:yeq_eta}-\ref{fig:yeq_components}).
\newline

\noindent
3. -- The outflow velocity is sensitive to the accuracy with which the nuclear binding energy
release is treated. Including a representative heavy nucleus (${}^{54}$Mn) results in an
additional energy release of $\sim 1$\,MeV per baryon relative to including only alpha particle
recombination (Figure~\ref{fig:nse}). Additional energy release can be obtained when including a much larger
isotope mixture. This motivates further work toward including more realistic nuclear physics
in post-merger simulations.
\newline

\noindent
4. -- The nucleosynthesis yields follow the offset in $Y_e$, with ALCAR models producing more 
elements with $A>130$ by a factor $\sim 2$ relative to FLASH models, within an overall good agreement otherwise. This also results in a higher heating rate in ALCAR models due to enhancement of the $^{132}\rm Te$--$^{132}\rm I$--$^{132}\rm Xe$ $\beta^-$ decay chain. Very minor differences result from the additional energy release from ${}^{54}$Mn models in FLASH (Fh versus F), with the possible exception of factor $\sim 2$ changes in the lanthanides fraction in some models.
Small differences in the amount of ejecta with $Y_e\sim 0.1$ can have a factor $\sim 10$ imprint in the abundance
of actinides, with the potential to use these species as a diagnostic of the electron fraction of the disk outflow.
\newline

\noindent
5. -- The kilonova signature is quite similar in all models after the ejecta become optically thin, 
despite the aforementioned differences in the heating rate. More pronounced differences are found at early times when the ejecta are optically thick, with ALCAR models being brighter due in part to a more extended high-velocity tail given the more prominent neutrino-driven wind. We also find evidence of undersampling of the early ejecta with $\sim 2000$ total particles, showing a significant brightening in FLASH models when $10$ times more particles are used.
\newline

The expected boost in expansion velocity when including a large number of 
isotopes ($\sim 0.1$\,c relative to pure alpha particles, Fig.~\ref{fig:nse})
is consistent with what has been found in studies looking at the impact of $r$-process
heating on the late-time evolution of disk outflows (e.g., \cite{klion_2022}).
Development of EOSs with more complete nuclear heating rates that also
cover the relevant thermodynamic range for post-merger evolution would be
of high usefulness to post-merger modeling.

Further code comparison studies are needed to bracket uncertainties in theoretical predictions
for kilonova light curves and $r$-process nucleosynthesis yields, as more events with
electromagnetic counterparts are anticipated in the future. While a first code comparison of GRMHD
models has been carried out recently covering short evolutionary timescales of tens of milliseconds \cite{espino_2022}, more extensive comparisons of GRMHD models include a microphysical EOS and neutrino transport are needed. The high computational costs of these calculations makes extensive
comparisons impractical at present, however, but nevertheless highly desirable for the future.

\appendix

\begin{acknowledgments}
RF acknowledges support from the Natural Sciences and Engineering Research Council of Canada (NSERC) through Discovery Grants RGPIN-2017-04286 and RGPIN-2022-03463. A sabbatical visit, during which this work was conceived and partially completed, was supported by the Cluster Project ELEMENTS from the State of Hesse and the European Research Council (ERC) under the European Union's Horizon 2020 research and innovation programme (ERC Advanced Grant KILONOVA No.~885281). RF is also grateful for
the hospitality of the GSI Helmholtz Centre for Heavy Ion Research and the Institute of Physics, Academia Sinica, where part of this work was also conducted. OJ acknowledges support by the ERC under the European Union's Horizon 2020 research and innovation programme under grant agreement nr 759253. GMP and ZX acknowledge support by the ERC under the European Union's Horizon 2020 research and innovation programme (ERC Advanced Grant KILONOVA No.~885281). 
OJ, GMP, and ZX also acknowledge support by Deutsche Forschungsgemeinschaft (DFG, German Research Foundation) - Project-ID 279384907 - SFB 1245 and MA 4248/3-1, and the State of Hesse within the Cluster Project ELEMENTS.
Some of the software used in this work was in part developed by U.S. Department of Energy (DOE) NNSA-ASC OASCR Flash Center at the University of Chicago. We also acknowledge support from the Shared Hierarchical Academic Research Computing Network (SHARCNET, www.sharcnet.ca) and the Digital Research Alliance of Canada (alliancecan.ca). FLASH models were run on the \emph{Niagara} supercomputer at the SciNet HPC Consortium \cite{SciNet,Niagara} and analyzed on the \emph{graham} cluster at the University of Waterloo. SciNet is funded by the Canada Foundation for Innovation, the Government of Ontario 
(Ontario Research Fund - Research
Excellence), and by the University of Toronto. 
RF also acknowledges storage resources of the National Energy Research Scientific Computing Center (NERSC), a DOE Office of Science User Facility using NERSC award NP-ERCAP0024213.
We are also grateful for computational support by the VIRGO cluster at GSI and the HOKUSAI computer center at RIKEN. \newline
\end{acknowledgments}

\section{Nuclear statistical equilibrium for ions \label{s:nse_appendix}}

Here we provide the explicit relations defining nuclear statistical
equilibrium for the ion component of the equation of state. Chemical equilibrium
leads to the stoichiometric equations for the chemical potentials
associated with reactions relating particle species, combined with
mass and charge conservation. For a gas of neutrons, protons,
alpha particles, and ${}^{54}$Mn nuclei, we have
\begin{eqnarray}
\label{eq:alpha_dissoc}
2\mu_{\rm n} + 2\mu_{\rm p} & = & \mu_\alpha\\
\label{eq:mn_dissoc}
12\mu_\alpha + 5\mu_{\rm n} + \mu_{\rm p} & = & \mu_{\rm Mn}\\
\label{eq:nse_mass_conservation}
X_{\rm n} + X_{\rm p} + X_\alpha + X_{\rm Mn} & = & 1\\
\label{eq:nse_charge_conservation}
X_{\rm p} + \frac{1}{2}X_\alpha + \frac{25}{54}X_{\rm Mn} & = & Y_e\\
& & \nonumber
\end{eqnarray}
where the subscripts \{n,p,$\alpha$,Mn\} correspond to neutrons,
protons, alpha particles, and ${}^{54}$Mn nuclei, $\mu_i$ are the
chemical potentials, $X_i$ are the mass fractions, and $Y_e$ is the
electron fraction. If particles follow a Maxwell-Boltzmann distribution,
we have 
\begin{equation}
\mu_i = k_{\rm B}T\left[\ln\left(\frac{n_i}{n_{\rm Q,i}}\right) - \ln \omega_i\right] - \chi_i,
\end{equation}
where $\chi_i$ is the nuclear binding energy, $\omega_i$ is the nuclear
partition function, $n_i$ is the number density, and
\begin{equation}
n_{\rm Q,i} = \left(\frac{m_i k_{\rm B}T}{2\pi \hbar^2} \right)^{3/2}
\end{equation}
is the quantum concentration of particle species $i$. The stoichiometric equations for the chemical
potential then become Saha equations for each dissociation/recombination channel.

Solution of the system of Equations (\ref{eq:alpha_dissoc})-(\ref{eq:nse_charge_conservation})
yields the equilibrium mass fractions for each species, as a function of density, temperature,
and electron fraction $X^{\rm NSE}_i(\rho,T,Y_e)$. The nuclear binding energy
contribution to the specific internal energy is included as
\begin{equation}
\label{eq:eint_nuclear_binding}
e_{\rm int} = e_{\rm int}^0 -\frac{\chi_\alpha}{m_\alpha}\,X_\alpha - \frac{\chi_{\rm Mn}}{m_{\rm Mn}}\,X_{\rm Mn} 
\end{equation}
where $e_{\rm int}^0$ is the internal energy excluding nuclear binding energy, 
and $m_i$ is the mass of particle species $i$. For a non-relativistic ion gas, we have
\begin{equation}
\label{eq:eint_kinetic_non-rel}
e_{\rm int}^0 = \frac{3}{2}\frac{k_{\rm B}T}{m_n}\sum_i \frac{X_i}{A_i}. 
\end{equation}
For an equation of state in which the temperature is found from the internal energy using a Newton-Raphson scheme
(as in FLASH), the derivatives of the mass fractions in NSE with respect to temperature must also be calculated
to obtain the total temperature derivative of the internal energy as defined in Equation~(\ref{eq:eint_nuclear_binding}).

The original FLASH implementation does not account for ${}^{54}$Mn nuclei, thus Equation~(\ref{eq:mn_dissoc}) is not
included, and $X_{\rm Mn}=0$ everywhere else (including in Equation~\ref{eq:eint_nuclear_binding}). NSE is solved
for by solving the Saha equation for $\alpha$ particles directly. To include ${}^{54}$Mn nuclei in the extended
set of simulations, we generate a table of \{n,p,$\alpha$,Mn\} mass fractions and associated temperature
derivatives $(\partial X^{\rm NSE}_i / \partial T)_{\rho,Y_e}$ using the code\footnote{Available at cococubed.com} 
of \cite{seitenzahl_2008}  and constant
partition functions $\omega_{\rm n}=\omega_{\rm p}=2$ and $\omega_{\alpha}=\omega_{\rm Mn}=1$. The table covers the 
range $T\in [5,100]\times 10^{9}$\,K, 
$\log_{\rm 10}\rho \in [1,12]$, and $Y_e \in [0.01,0.99]$. The ion internal energy is then computed 
using Equations~(\ref{eq:eint_nuclear_binding}) and (\ref{eq:eint_kinetic_non-rel}).

\bibliography{oliver,rodrigo}

\begin{thebibliography}{78}%
\makeatletter
\providecommand \@ifxundefined [1]{%
 \@ifx{#1\undefined}
}%
\providecommand \@ifnum [1]{%
 \ifnum #1\expandafter \@firstoftwo
 \else \expandafter \@secondoftwo
 \fi
}%
\providecommand \@ifx [1]{%
 \ifx #1\expandafter \@firstoftwo
 \else \expandafter \@secondoftwo
 \fi
}%
\providecommand \natexlab [1]{#1}%
\providecommand \enquote  [1]{``#1''}%
\providecommand \bibnamefont  [1]{#1}%
\providecommand \bibfnamefont [1]{#1}%
\providecommand \citenamefont [1]{#1}%
\providecommand \href@noop [0]{\@secondoftwo}%
\providecommand \href [0]{\begingroup \@sanitize@url \@href}%
\providecommand \@href[1]{\@@startlink{#1}\@@href}%
\providecommand \@@href[1]{\endgroup#1\@@endlink}%
\providecommand \@sanitize@url [0]{\catcode `\\12\catcode `\$12\catcode
  `\&12\catcode `\#12\catcode `\^12\catcode `\_12\catcode `\%12\relax}%
\providecommand \@@startlink[1]{}%
\providecommand \@@endlink[0]{}%
\providecommand \url  [0]{\begingroup\@sanitize@url \@url }%
\providecommand \@url [1]{\endgroup\@href {#1}{\urlprefix }}%
\providecommand \urlprefix  [0]{URL }%
\providecommand \Eprint [0]{\href }%
\providecommand \doibase [0]{https://doi.org/}%
\providecommand \selectlanguage [0]{\@gobble}%
\providecommand \bibinfo  [0]{\@secondoftwo}%
\providecommand \bibfield  [0]{\@secondoftwo}%
\providecommand \translation [1]{[#1]}%
\providecommand \BibitemOpen [0]{}%
\providecommand \bibitemStop [0]{}%
\providecommand \bibitemNoStop [0]{.\EOS\space}%
\providecommand \EOS [0]{\spacefactor3000\relax}%
\providecommand \BibitemShut  [1]{\csname bibitem#1\endcsname}%
\let\auto@bib@innerbib\@empty
\bibitem [{\citenamefont {{Abbott}}\ \emph {et~al.}(2017)\citenamefont
  {{Abbott}} \emph {et~al.}}]{ligo_gw170817_multi-messenger}%
  \BibitemOpen
  \bibfield  {author} {\bibinfo {author} {\bibfnamefont {B.~P.}\ \bibnamefont
  {{Abbott}}} \emph {et~al.},\ }\bibfield  {title} {\bibinfo {title}
  {{Multi-messenger Observations of a Binary Neutron Star Merger}},\ }\href
  {https://doi.org/10.3847/2041-8213/aa91c9} {\bibfield  {journal} {\bibinfo
  {journal} {ApJ}\ }\textbf {\bibinfo {volume} {848}},\ \bibinfo {eid} {L12}
  (\bibinfo {year} {2017})},\ \Eprint {https://arxiv.org/abs/1710.05833}
  {arXiv:1710.05833 [astro-ph.HE]} \BibitemShut {NoStop}%
\bibitem [{\citenamefont {{Drout}}\ \emph {et~al.}(2017)\citenamefont {{Drout}}
  \emph {et~al.}}]{drout_2017}%
  \BibitemOpen
  \bibfield  {author} {\bibinfo {author} {\bibfnamefont {M.~R.}\ \bibnamefont
  {{Drout}}} \emph {et~al.},\ }\bibfield  {title} {\bibinfo {title} {{Light
  curves of the neutron star merger GW170817/SSS17a: Implications for r-process
  nucleosynthesis}},\ }\href {https://doi.org/10.1126/science.aaq0049}
  {\bibfield  {journal} {\bibinfo  {journal} {Science}\ }\textbf {\bibinfo
  {volume} {358}},\ \bibinfo {pages} {1570} (\bibinfo {year} {2017})},\ \Eprint
  {https://arxiv.org/abs/1710.05443} {arXiv:1710.05443 [astro-ph.HE]}
  \BibitemShut {NoStop}%
\bibitem [{\citenamefont {{Fern{\'a}ndez}}\ and\ \citenamefont
  {{Metzger}}(2016)}]{FM16}%
  \BibitemOpen
  \bibfield  {author} {\bibinfo {author} {\bibfnamefont {R.}~\bibnamefont
  {{Fern{\'a}ndez}}}\ and\ \bibinfo {author} {\bibfnamefont {B.~D.}\
  \bibnamefont {{Metzger}}},\ }\bibfield  {title} {\bibinfo {title}
  {{Electromagnetic Signatures of Neutron Star Mergers in the Advanced LIGO
  Era}},\ }\href {https://doi.org/10.1146/annurev-nucl-102115-044819}
  {\bibfield  {journal} {\bibinfo  {journal} {ARNPS}\ }\textbf {\bibinfo
  {volume} {66}},\ \bibinfo {pages} {23} (\bibinfo {year} {2016})},\ \Eprint
  {https://arxiv.org/abs/1512.05435} {arXiv:1512.05435 [astro-ph.HE]}
  \BibitemShut {NoStop}%
\bibitem [{\citenamefont {Baiotti}\ and\ \citenamefont
  {Rezzolla}(2017)}]{baiotti_BinaryNeutronStar_2017}%
  \BibitemOpen
  \bibfield  {author} {\bibinfo {author} {\bibfnamefont {L.}~\bibnamefont
  {Baiotti}}\ and\ \bibinfo {author} {\bibfnamefont {L.}~\bibnamefont
  {Rezzolla}},\ }\bibfield  {title} {\bibinfo {title} {Binary neutron star
  mergers: a review of {Einstein}ârichest laboratory},\ }\href
  {https://doi.org/10.1088/1361-6633/aa67bb} {\bibfield  {journal} {\bibinfo
  {journal} {Reports on Progress in Physics}\ }\textbf {\bibinfo {volume}
  {80}},\ \bibinfo {pages} {096901} (\bibinfo {year} {2017})}\BibitemShut
  {NoStop}%
\bibitem [{\citenamefont {Radice}\ \emph {et~al.}(2020)\citenamefont {Radice},
  \citenamefont {Bernuzzi},\ and\ \citenamefont
  {Perego}}]{radice_DynamicsBinaryNeutron_2020}%
  \BibitemOpen
  \bibfield  {author} {\bibinfo {author} {\bibfnamefont {D.}~\bibnamefont
  {Radice}}, \bibinfo {author} {\bibfnamefont {S.}~\bibnamefont {Bernuzzi}},\
  and\ \bibinfo {author} {\bibfnamefont {A.}~\bibnamefont {Perego}},\
  }\bibfield  {title} {\bibinfo {title} {The {Dynamics} of {Binary} {Neutron}
  {Star} {Mergers} and {GW170817}},\ }\href
  {https://doi.org/10.1146/annurev-nucl-013120-114541} {\bibfield  {journal}
  {\bibinfo  {journal} {Annual Review of Nuclear and Particle Science}\
  }\textbf {\bibinfo {volume} {70}},\ \bibinfo {pages} {95} (\bibinfo {year}
  {2020})}\BibitemShut {NoStop}%
\bibitem [{\citenamefont {{Janka}}\ and\ \citenamefont
  {{Bauswein}}(2022)}]{janka_bauswein_2022_review}%
  \BibitemOpen
  \bibfield  {author} {\bibinfo {author} {\bibfnamefont {H.~T.}\ \bibnamefont
  {{Janka}}}\ and\ \bibinfo {author} {\bibfnamefont {A.}~\bibnamefont
  {{Bauswein}}},\ }\bibfield  {title} {\bibinfo {title} {{Dynamics and Equation
  of State Dependencies of Relevance for Nucleosynthesis in Supernovae and
  Neutron Star Mergers}},\ }in\ \href@noop {} {\emph {\bibinfo {booktitle}
  {Handbook of Nuclear Physics}}},\ \bibinfo {series} {TBD Series},
  Vol.~\bibinfo {volume} {1}\ (\bibinfo {year} {2022})\ p.\ \bibinfo {pages}
  {arXiv:2212.07498},\ \Eprint {https://arxiv.org/abs/2212.07498}
  {arXiv:2212.07498 [astro-ph.HE]} \BibitemShut {NoStop}%
\bibitem [{\citenamefont {{Siegel}}\ and\ \citenamefont
  {{Metzger}}(2018)}]{siegel_2018}%
  \BibitemOpen
  \bibfield  {author} {\bibinfo {author} {\bibfnamefont {D.~M.}\ \bibnamefont
  {{Siegel}}}\ and\ \bibinfo {author} {\bibfnamefont {B.~D.}\ \bibnamefont
  {{Metzger}}},\ }\bibfield  {title} {\bibinfo {title} {{Three-dimensional
  GRMHD Simulations of Neutrino-cooled Accretion Disks from Neutron Star
  Mergers}},\ }\href {https://doi.org/10.3847/1538-4357/aabaec} {\bibfield
  {journal} {\bibinfo  {journal} {ApJ}\ }\textbf {\bibinfo {volume} {858}},\
  \bibinfo {eid} {52} (\bibinfo {year} {2018})},\ \Eprint
  {https://arxiv.org/abs/1711.00868} {arXiv:1711.00868 [astro-ph.HE]}
  \BibitemShut {NoStop}%
\bibitem [{\citenamefont {{Just}}\ \emph
  {et~al.}(2022{\natexlab{a}})\citenamefont {{Just}}, \citenamefont
  {{Goriely}}, \citenamefont {{Janka}}, \citenamefont {{Nagataki}},\ and\
  \citenamefont {{Bauswein}}}]{Just2022_Yeq}%
  \BibitemOpen
  \bibfield  {author} {\bibinfo {author} {\bibfnamefont {O.}~\bibnamefont
  {{Just}}}, \bibinfo {author} {\bibfnamefont {S.}~\bibnamefont {{Goriely}}},
  \bibinfo {author} {\bibfnamefont {H.~T.}\ \bibnamefont {{Janka}}}, \bibinfo
  {author} {\bibfnamefont {S.}~\bibnamefont {{Nagataki}}},\ and\ \bibinfo
  {author} {\bibfnamefont {A.}~\bibnamefont {{Bauswein}}},\ }\bibfield  {title}
  {\bibinfo {title} {{Neutrino absorption and other physics dependencies in
  neutrino-cooled black hole accretion discs}},\ }\href
  {https://doi.org/10.1093/mnras/stab2861} {\bibfield  {journal} {\bibinfo
  {journal} {\mnras}\ }\textbf {\bibinfo {volume} {509}},\ \bibinfo {pages}
  {1377} (\bibinfo {year} {2022}{\natexlab{a}})},\ \Eprint
  {https://arxiv.org/abs/2102.08387} {arXiv:2102.08387 [astro-ph.HE]}
  \BibitemShut {NoStop}%
\bibitem [{\citenamefont {{Popham}}\ \emph {et~al.}(1999)\citenamefont
  {{Popham}}, \citenamefont {{Woosley}},\ and\ \citenamefont
  {{Fryer}}}]{popham1999}%
  \BibitemOpen
  \bibfield  {author} {\bibinfo {author} {\bibfnamefont {R.}~\bibnamefont
  {{Popham}}}, \bibinfo {author} {\bibfnamefont {S.~E.}\ \bibnamefont
  {{Woosley}}},\ and\ \bibinfo {author} {\bibfnamefont {C.}~\bibnamefont
  {{Fryer}}},\ }\bibfield  {title} {\bibinfo {title} {{Hyperaccreting Black
  Holes and Gamma-Ray Bursts}},\ }\href@noop {} {\bibfield  {journal} {\bibinfo
   {journal} {ApJ}\ }\textbf {\bibinfo {volume} {518}},\ \bibinfo {pages} {356}
  (\bibinfo {year} {1999})}\BibitemShut {NoStop}%
\bibitem [{\citenamefont {{Di Matteo}}\ \emph {et~al.}(2002)\citenamefont {{Di
  Matteo}}, \citenamefont {{Perna}},\ and\ \citenamefont
  {{Narayan}}}]{DiMatteo+02}%
  \BibitemOpen
  \bibfield  {author} {\bibinfo {author} {\bibfnamefont {T.}~\bibnamefont {{Di
  Matteo}}}, \bibinfo {author} {\bibfnamefont {R.}~\bibnamefont {{Perna}}},\
  and\ \bibinfo {author} {\bibfnamefont {R.}~\bibnamefont {{Narayan}}},\
  }\bibfield  {title} {\bibinfo {title} {{Neutrino Trapping and Accretion
  Models for Gamma-Ray Bursts}},\ }\href {https://doi.org/10.1086/342832}
  {\bibfield  {journal} {\bibinfo  {journal} {ApJ}\ }\textbf {\bibinfo {volume}
  {579}},\ \bibinfo {pages} {706} (\bibinfo {year} {2002})},\ \Eprint
  {https://arxiv.org/abs/arXiv:astro-ph/0207319} {arXiv:astro-ph/0207319}
  \BibitemShut {NoStop}%
\bibitem [{\citenamefont {{Setiawan}}\ \emph {et~al.}(2004)\citenamefont
  {{Setiawan}}, \citenamefont {{Ruffert}},\ and\ \citenamefont
  {{Janka}}}]{setiawan2004}%
  \BibitemOpen
  \bibfield  {author} {\bibinfo {author} {\bibfnamefont {S.}~\bibnamefont
  {{Setiawan}}}, \bibinfo {author} {\bibfnamefont {M.}~\bibnamefont
  {{Ruffert}}},\ and\ \bibinfo {author} {\bibfnamefont {H.-T.}\ \bibnamefont
  {{Janka}}},\ }\bibfield  {title} {\bibinfo {title} {{Non-stationary
  hyperaccretion of stellar-mass black holes in three dimensions: torus
  evolution and neutrino emission}},\ }\href@noop {} {\bibfield  {journal}
  {\bibinfo  {journal} {MNRAS}\ }\textbf {\bibinfo {volume} {352}},\ \bibinfo
  {pages} {753} (\bibinfo {year} {2004})}\BibitemShut {NoStop}%
\bibitem [{\citenamefont {{Chen}}\ and\ \citenamefont
  {{Beloborodov}}(2007)}]{Chen&Beloborodov07}%
  \BibitemOpen
  \bibfield  {author} {\bibinfo {author} {\bibfnamefont {W.-X.}\ \bibnamefont
  {{Chen}}}\ and\ \bibinfo {author} {\bibfnamefont {A.~M.}\ \bibnamefont
  {{Beloborodov}}},\ }\bibfield  {title} {\bibinfo {title} {{Neutrino-cooled
  Accretion Disks around Spinning Black Holes}},\ }\href
  {https://doi.org/10.1086/508923} {\bibfield  {journal} {\bibinfo  {journal}
  {ApJ}\ }\textbf {\bibinfo {volume} {657}},\ \bibinfo {pages} {383} (\bibinfo
  {year} {2007})},\ \Eprint {https://arxiv.org/abs/arXiv:astro-ph/0607145}
  {arXiv:astro-ph/0607145} \BibitemShut {NoStop}%
\bibitem [{\citenamefont {{De}}\ and\ \citenamefont
  {{Siegel}}(2021)}]{de_siegel_2021}%
  \BibitemOpen
  \bibfield  {author} {\bibinfo {author} {\bibfnamefont {S.}~\bibnamefont
  {{De}}}\ and\ \bibinfo {author} {\bibfnamefont {D.~M.}\ \bibnamefont
  {{Siegel}}},\ }\bibfield  {title} {\bibinfo {title} {{Igniting Weak
  Interactions in Neutron Star Postmerger Accretion Disks}},\ }\href
  {https://doi.org/10.3847/1538-4357/ac110b} {\bibfield  {journal} {\bibinfo
  {journal} {\apj}\ }\textbf {\bibinfo {volume} {921}},\ \bibinfo {eid} {94}
  (\bibinfo {year} {2021})},\ \Eprint {https://arxiv.org/abs/2011.07176}
  {arXiv:2011.07176 [astro-ph.HE]} \BibitemShut {NoStop}%
\bibitem [{\citenamefont {{Metzger}}\ \emph {et~al.}(2009)\citenamefont
  {{Metzger}}, \citenamefont {{Piro}},\ and\ \citenamefont
  {{Quataert}}}]{Metzger+09a}%
  \BibitemOpen
  \bibfield  {author} {\bibinfo {author} {\bibfnamefont {B.~D.}\ \bibnamefont
  {{Metzger}}}, \bibinfo {author} {\bibfnamefont {A.~L.}\ \bibnamefont
  {{Piro}}},\ and\ \bibinfo {author} {\bibfnamefont {E.}~\bibnamefont
  {{Quataert}}},\ }\bibfield  {title} {\bibinfo {title} {{Neutron-rich
  freeze-out in viscously spreading accretion discs formed from compact object
  mergers}},\ }\href {https://doi.org/10.1111/j.1365-2966.2008.14380.x}
  {\bibfield  {journal} {\bibinfo  {journal} {MNRAS}\ }\textbf {\bibinfo
  {volume} {396}},\ \bibinfo {pages} {304} (\bibinfo {year} {2009})},\ \Eprint
  {https://arxiv.org/abs/0810.2535} {arXiv:0810.2535} \BibitemShut {NoStop}%
\bibitem [{\citenamefont {{Metzger}}\ and\ \citenamefont
  {{Fern{\'a}ndez}}(2014)}]{MF14}%
  \BibitemOpen
  \bibfield  {author} {\bibinfo {author} {\bibfnamefont {B.~D.}\ \bibnamefont
  {{Metzger}}}\ and\ \bibinfo {author} {\bibfnamefont {R.}~\bibnamefont
  {{Fern{\'a}ndez}}},\ }\bibfield  {title} {\bibinfo {title} {{Red or blue? A
  potential kilonova imprint of the delay until black hole formation following
  a neutron star merger}},\ }\href {https://doi.org/10.1093/mnras/stu802}
  {\bibfield  {journal} {\bibinfo  {journal} {\mnras}\ }\textbf {\bibinfo
  {volume} {441}},\ \bibinfo {pages} {3444} (\bibinfo {year} {2014})},\ \Eprint
  {https://arxiv.org/abs/1402.4803} {arXiv:1402.4803 [astro-ph.HE]}
  \BibitemShut {NoStop}%
\bibitem [{\citenamefont {{Just}}\ \emph
  {et~al.}(2015{\natexlab{a}})\citenamefont {{Just}}, \citenamefont
  {{Bauswein}}, \citenamefont {{Pulpillo}}, \citenamefont {{Goriely}},\ and\
  \citenamefont {{Janka}}}]{Just2015a}%
  \BibitemOpen
  \bibfield  {author} {\bibinfo {author} {\bibfnamefont {O.}~\bibnamefont
  {{Just}}}, \bibinfo {author} {\bibfnamefont {A.}~\bibnamefont {{Bauswein}}},
  \bibinfo {author} {\bibfnamefont {R.~A.}\ \bibnamefont {{Pulpillo}}},
  \bibinfo {author} {\bibfnamefont {S.}~\bibnamefont {{Goriely}}},\ and\
  \bibinfo {author} {\bibfnamefont {H.-T.}\ \bibnamefont {{Janka}}},\
  }\bibfield  {title} {\bibinfo {title} {{Comprehensive nucleosynthesis
  analysis for ejecta of compact binary mergers}},\ }\href@noop {} {\bibfield
  {journal} {\bibinfo  {journal} {\mnras}\ }\textbf {\bibinfo {volume} {448}},\
  \bibinfo {pages} {541} (\bibinfo {year} {2015}{\natexlab{a}})}\BibitemShut
  {NoStop}%
\bibitem [{\citenamefont {{Fujibayashi}}\ \emph {et~al.}(2018)\citenamefont
  {{Fujibayashi}}, \citenamefont {{Kiuchi}}, \citenamefont {{Nishimura}},
  \citenamefont {{Sekiguchi}},\ and\ \citenamefont
  {{Shibata}}}]{fujibayashi_2018}%
  \BibitemOpen
  \bibfield  {author} {\bibinfo {author} {\bibfnamefont {S.}~\bibnamefont
  {{Fujibayashi}}}, \bibinfo {author} {\bibfnamefont {K.}~\bibnamefont
  {{Kiuchi}}}, \bibinfo {author} {\bibfnamefont {N.}~\bibnamefont
  {{Nishimura}}}, \bibinfo {author} {\bibfnamefont {Y.}~\bibnamefont
  {{Sekiguchi}}},\ and\ \bibinfo {author} {\bibfnamefont {M.}~\bibnamefont
  {{Shibata}}},\ }\bibfield  {title} {\bibinfo {title} {{Mass Ejection from the
  Remnant of a Binary Neutron Star Merger: Viscous-radiation Hydrodynamics
  Study}},\ }\href {https://doi.org/10.3847/1538-4357/aabafd} {\bibfield
  {journal} {\bibinfo  {journal} {ApJ}\ }\textbf {\bibinfo {volume} {860}},\
  \bibinfo {eid} {64} (\bibinfo {year} {2018})},\ \Eprint
  {https://arxiv.org/abs/1711.02093} {arXiv:1711.02093 [astro-ph.HE]}
  \BibitemShut {NoStop}%
\bibitem [{\citenamefont {{Fujibayashi}}\ \emph
  {et~al.}(2020{\natexlab{a}})\citenamefont {{Fujibayashi}}, \citenamefont
  {{Wanajo}}, \citenamefont {{Kiuchi}}, \citenamefont {{Kyutoku}},
  \citenamefont {{Sekiguchi}},\ and\ \citenamefont
  {{Shibata}}}]{fujibayashi_2020_ns}%
  \BibitemOpen
  \bibfield  {author} {\bibinfo {author} {\bibfnamefont {S.}~\bibnamefont
  {{Fujibayashi}}}, \bibinfo {author} {\bibfnamefont {S.}~\bibnamefont
  {{Wanajo}}}, \bibinfo {author} {\bibfnamefont {K.}~\bibnamefont {{Kiuchi}}},
  \bibinfo {author} {\bibfnamefont {K.}~\bibnamefont {{Kyutoku}}}, \bibinfo
  {author} {\bibfnamefont {Y.}~\bibnamefont {{Sekiguchi}}},\ and\ \bibinfo
  {author} {\bibfnamefont {M.}~\bibnamefont {{Shibata}}},\ }\bibfield  {title}
  {\bibinfo {title} {{Postmerger Mass Ejection of Low-mass Binary Neutron
  Stars}},\ }\href {https://doi.org/10.3847/1538-4357/abafc2} {\bibfield
  {journal} {\bibinfo  {journal} {\apj}\ }\textbf {\bibinfo {volume} {901}},\
  \bibinfo {eid} {122} (\bibinfo {year} {2020}{\natexlab{a}})},\ \Eprint
  {https://arxiv.org/abs/2007.00474} {arXiv:2007.00474 [astro-ph.HE]}
  \BibitemShut {NoStop}%
\bibitem [{\citenamefont {{Christie}}\ \emph {et~al.}(2019)\citenamefont
  {{Christie}}, \citenamefont {{Lalakos}}, \citenamefont {{Tchekhovskoy}},
  \citenamefont {{Fern{\'a}ndez}}, \citenamefont {{Foucart}}, \citenamefont
  {{Quataert}},\ and\ \citenamefont {{Kasen}}}]{christie2019}%
  \BibitemOpen
  \bibfield  {author} {\bibinfo {author} {\bibfnamefont {I.~M.}\ \bibnamefont
  {{Christie}}}, \bibinfo {author} {\bibfnamefont {A.}~\bibnamefont
  {{Lalakos}}}, \bibinfo {author} {\bibfnamefont {A.}~\bibnamefont
  {{Tchekhovskoy}}}, \bibinfo {author} {\bibfnamefont {R.}~\bibnamefont
  {{Fern{\'a}ndez}}}, \bibinfo {author} {\bibfnamefont {F.}~\bibnamefont
  {{Foucart}}}, \bibinfo {author} {\bibfnamefont {E.}~\bibnamefont
  {{Quataert}}},\ and\ \bibinfo {author} {\bibfnamefont {D.}~\bibnamefont
  {{Kasen}}},\ }\bibfield  {title} {\bibinfo {title} {{The role of magnetic
  field geometry in the evolution of neutron star merger accretion discs}},\
  }\href {https://doi.org/10.1093/mnras/stz2552} {\bibfield  {journal}
  {\bibinfo  {journal} {\mnras}\ }\textbf {\bibinfo {volume} {490}},\ \bibinfo
  {pages} {4811} (\bibinfo {year} {2019})},\ \Eprint
  {https://arxiv.org/abs/1907.02079} {arXiv:1907.02079 [astro-ph.HE]}
  \BibitemShut {NoStop}%
\bibitem [{\citenamefont {{Hayashi}}\ \emph
  {et~al.}(2022{\natexlab{a}})\citenamefont {{Hayashi}}, \citenamefont
  {{Fujibayashi}}, \citenamefont {{Kiuchi}}, \citenamefont {{Kyutoku}},
  \citenamefont {{Sekiguchi}},\ and\ \citenamefont
  {{Shibata}}}]{hayashi_2022a}%
  \BibitemOpen
  \bibfield  {author} {\bibinfo {author} {\bibfnamefont {K.}~\bibnamefont
  {{Hayashi}}}, \bibinfo {author} {\bibfnamefont {S.}~\bibnamefont
  {{Fujibayashi}}}, \bibinfo {author} {\bibfnamefont {K.}~\bibnamefont
  {{Kiuchi}}}, \bibinfo {author} {\bibfnamefont {K.}~\bibnamefont {{Kyutoku}}},
  \bibinfo {author} {\bibfnamefont {Y.}~\bibnamefont {{Sekiguchi}}},\ and\
  \bibinfo {author} {\bibfnamefont {M.}~\bibnamefont {{Shibata}}},\ }\bibfield
  {title} {\bibinfo {title} {{General-relativistic neutrino-radiation
  magnetohydrodynamic simulation of seconds-long black hole-neutron star
  mergers}},\ }\href {https://doi.org/10.1103/PhysRevD.106.023008} {\bibfield
  {journal} {\bibinfo  {journal} {\prd}\ }\textbf {\bibinfo {volume} {106}},\
  \bibinfo {eid} {023008} (\bibinfo {year} {2022}{\natexlab{a}})}\BibitemShut
  {NoStop}%
\bibitem [{\citenamefont {{Hayashi}}\ \emph
  {et~al.}(2022{\natexlab{b}})\citenamefont {{Hayashi}}, \citenamefont
  {{Kiuchi}}, \citenamefont {{Kyutoku}}, \citenamefont {{Sekiguchi}},\ and\
  \citenamefont {{Shibata}}}]{hayashi_2022b}%
  \BibitemOpen
  \bibfield  {author} {\bibinfo {author} {\bibfnamefont {K.}~\bibnamefont
  {{Hayashi}}}, \bibinfo {author} {\bibfnamefont {K.}~\bibnamefont {{Kiuchi}}},
  \bibinfo {author} {\bibfnamefont {K.}~\bibnamefont {{Kyutoku}}}, \bibinfo
  {author} {\bibfnamefont {Y.}~\bibnamefont {{Sekiguchi}}},\ and\ \bibinfo
  {author} {\bibfnamefont {M.}~\bibnamefont {{Shibata}}},\ }\bibfield  {title}
  {\bibinfo {title} {{General-relativistic neutrino-radiation
  magnetohydrodynamics simulation of seconds-long black hole-neutron star
  mergers: Dependence on initial magnetic field strength, configuration, and
  neutron-star equation of state}},\ }\href@noop {} {\bibfield  {journal}
  {\bibinfo  {journal} {arXiv e-prints}\ ,\ \bibinfo {eid} {arXiv:2211.07158}}
  (\bibinfo {year} {2022}{\natexlab{b}})},\ \Eprint
  {https://arxiv.org/abs/2211.07158} {arXiv:2211.07158 [astro-ph.HE]}
  \BibitemShut {NoStop}%
\bibitem [{\citenamefont {{Fern{\'a}ndez}}\ and\ \citenamefont
  {{Metzger}}(2013{\natexlab{a}})}]{FM13}%
  \BibitemOpen
  \bibfield  {author} {\bibinfo {author} {\bibfnamefont {R.}~\bibnamefont
  {{Fern{\'a}ndez}}}\ and\ \bibinfo {author} {\bibfnamefont {B.~D.}\
  \bibnamefont {{Metzger}}},\ }\bibfield  {title} {\bibinfo {title} {{Delayed
  outflows from black hole accretion tori following neutron star binary
  coalescence}},\ }\href {https://doi.org/10.1093/mnras/stt1312} {\bibfield
  {journal} {\bibinfo  {journal} {MNRAS}\ }\textbf {\bibinfo {volume} {435}},\
  \bibinfo {pages} {502} (\bibinfo {year} {2013}{\natexlab{a}})},\ \Eprint
  {https://arxiv.org/abs/1304.6720} {arXiv:1304.6720 [astro-ph.HE]}
  \BibitemShut {NoStop}%
\bibitem [{\citenamefont {{Fern{\'a}ndez}}\ \emph
  {et~al.}(2015{\natexlab{a}})\citenamefont {{Fern{\'a}ndez}}, \citenamefont
  {{Kasen}}, \citenamefont {{Metzger}},\ and\ \citenamefont
  {{Quataert}}}]{FKMQ14}%
  \BibitemOpen
  \bibfield  {author} {\bibinfo {author} {\bibfnamefont {R.}~\bibnamefont
  {{Fern{\'a}ndez}}}, \bibinfo {author} {\bibfnamefont {D.}~\bibnamefont
  {{Kasen}}}, \bibinfo {author} {\bibfnamefont {B.~D.}\ \bibnamefont
  {{Metzger}}},\ and\ \bibinfo {author} {\bibfnamefont {E.}~\bibnamefont
  {{Quataert}}},\ }\bibfield  {title} {\bibinfo {title} {{Outflows from
  accretion discs formed in neutron star mergers: effect of black hole spin}},\
  }\href {https://doi.org/10.1093/mnras/stu2112} {\bibfield  {journal}
  {\bibinfo  {journal} {MNRAS}\ }\textbf {\bibinfo {volume} {446}},\ \bibinfo
  {pages} {750} (\bibinfo {year} {2015}{\natexlab{a}})}\BibitemShut {NoStop}%
\bibitem [{\citenamefont {{Fern{\'a}ndez}}\ \emph
  {et~al.}(2015{\natexlab{b}})\citenamefont {{Fern{\'a}ndez}}, \citenamefont
  {{Quataert}}, \citenamefont {{Schwab}}, \citenamefont {{Kasen}},\ and\
  \citenamefont {{Rosswog}}}]{FQSKR-15}%
  \BibitemOpen
  \bibfield  {author} {\bibinfo {author} {\bibfnamefont {R.}~\bibnamefont
  {{Fern{\'a}ndez}}}, \bibinfo {author} {\bibfnamefont {E.}~\bibnamefont
  {{Quataert}}}, \bibinfo {author} {\bibfnamefont {J.}~\bibnamefont
  {{Schwab}}}, \bibinfo {author} {\bibfnamefont {D.}~\bibnamefont {{Kasen}}},\
  and\ \bibinfo {author} {\bibfnamefont {S.}~\bibnamefont {{Rosswog}}},\
  }\bibfield  {title} {\bibinfo {title} {{The interplay of disc wind and
  dynamical ejecta in the aftermath of neutron star-black hole mergers}},\
  }\href {https://doi.org/10.1093/mnras/stv238} {\bibfield  {journal} {\bibinfo
   {journal} {MNRAS}\ }\textbf {\bibinfo {volume} {449}},\ \bibinfo {pages}
  {390} (\bibinfo {year} {2015}{\natexlab{b}})},\ \Eprint
  {https://arxiv.org/abs/1412.5588} {arXiv:1412.5588 [astro-ph.HE]}
  \BibitemShut {NoStop}%
\bibitem [{\citenamefont {{Fern{\'a}ndez}}\ \emph {et~al.}(2017)\citenamefont
  {{Fern{\'a}ndez}}, \citenamefont {{Foucart}}, \citenamefont {{Kasen}},
  \citenamefont {{Lippuner}}, \citenamefont {{Desai}},\ and\ \citenamefont
  {{Roberts}}}]{fernandez_2017}%
  \BibitemOpen
  \bibfield  {author} {\bibinfo {author} {\bibfnamefont {R.}~\bibnamefont
  {{Fern{\'a}ndez}}}, \bibinfo {author} {\bibfnamefont {F.}~\bibnamefont
  {{Foucart}}}, \bibinfo {author} {\bibfnamefont {D.}~\bibnamefont {{Kasen}}},
  \bibinfo {author} {\bibfnamefont {J.}~\bibnamefont {{Lippuner}}}, \bibinfo
  {author} {\bibfnamefont {D.}~\bibnamefont {{Desai}}},\ and\ \bibinfo {author}
  {\bibfnamefont {L.~F.}\ \bibnamefont {{Roberts}}},\ }\bibfield  {title}
  {\bibinfo {title} {{Dynamics, nucleosynthesis, and kilonova signature of
  black hole--neutron star merger ejecta}},\ }\href
  {https://doi.org/10.1088/1361-6382/aa7a77} {\bibfield  {journal} {\bibinfo
  {journal} {CQG}\ }\textbf {\bibinfo {volume} {34}},\ \bibinfo {eid} {154001}
  (\bibinfo {year} {2017})},\ \Eprint {https://arxiv.org/abs/1612.04829}
  {arXiv:1612.04829 [astro-ph.HE]} \BibitemShut {NoStop}%
\bibitem [{\citenamefont {{Fahlman}}\ and\ \citenamefont
  {{Fern{\'a}ndez}}(2018)}]{fahlman_2018}%
  \BibitemOpen
  \bibfield  {author} {\bibinfo {author} {\bibfnamefont {S.}~\bibnamefont
  {{Fahlman}}}\ and\ \bibinfo {author} {\bibfnamefont {R.}~\bibnamefont
  {{Fern{\'a}ndez}}},\ }\bibfield  {title} {\bibinfo {title} {{Hypermassive
  Neutron Star Disk Outflows and Blue Kilonovae}},\ }\href
  {https://doi.org/10.3847/2041-8213/aaf1ab} {\bibfield  {journal} {\bibinfo
  {journal} {ApJ}\ }\textbf {\bibinfo {volume} {869}},\ \bibinfo {eid} {L3}
  (\bibinfo {year} {2018})},\ \Eprint {https://arxiv.org/abs/1811.08906}
  {arXiv:1811.08906 [astro-ph.HE]} \BibitemShut {NoStop}%
\bibitem [{\citenamefont {{Fujibayashi}}\ \emph
  {et~al.}(2020{\natexlab{b}})\citenamefont {{Fujibayashi}}, \citenamefont
  {{Shibata}}, \citenamefont {{Wanajo}}, \citenamefont {{Kiuchi}},
  \citenamefont {{Kyutoku}},\ and\ \citenamefont
  {{Sekiguchi}}}]{fujibayashi2020}%
  \BibitemOpen
  \bibfield  {author} {\bibinfo {author} {\bibfnamefont {S.}~\bibnamefont
  {{Fujibayashi}}}, \bibinfo {author} {\bibfnamefont {M.}~\bibnamefont
  {{Shibata}}}, \bibinfo {author} {\bibfnamefont {S.}~\bibnamefont {{Wanajo}}},
  \bibinfo {author} {\bibfnamefont {K.}~\bibnamefont {{Kiuchi}}}, \bibinfo
  {author} {\bibfnamefont {K.}~\bibnamefont {{Kyutoku}}},\ and\ \bibinfo
  {author} {\bibfnamefont {Y.}~\bibnamefont {{Sekiguchi}}},\ }\bibfield
  {title} {\bibinfo {title} {{Mass ejection from disks surrounding a low-mass
  black hole: Viscous neutrino-radiation hydrodynamics simulation in full
  general relativity}},\ }\href {https://doi.org/10.1103/PhysRevD.101.083029}
  {\bibfield  {journal} {\bibinfo  {journal} {PRD}\ }\textbf {\bibinfo {volume}
  {101}},\ \bibinfo {eid} {083029} (\bibinfo {year} {2020}{\natexlab{b}})},\
  \Eprint {https://arxiv.org/abs/2001.04467} {arXiv:2001.04467 [astro-ph.HE]}
  \BibitemShut {NoStop}%
\bibitem [{\citenamefont {{Fern{\'a}ndez}}\ \emph {et~al.}(2020)\citenamefont
  {{Fern{\'a}ndez}}, \citenamefont {{Foucart}},\ and\ \citenamefont
  {{Lippuner}}}]{Fernandez2020BHNS}%
  \BibitemOpen
  \bibfield  {author} {\bibinfo {author} {\bibfnamefont {R.}~\bibnamefont
  {{Fern{\'a}ndez}}}, \bibinfo {author} {\bibfnamefont {F.}~\bibnamefont
  {{Foucart}}},\ and\ \bibinfo {author} {\bibfnamefont {J.}~\bibnamefont
  {{Lippuner}}},\ }\bibfield  {title} {\bibinfo {title} {{The landscape of disc
  outflows from black hole-neutron star mergers}},\ }\href
  {https://doi.org/10.1093/mnras/staa2209} {\bibfield  {journal} {\bibinfo
  {journal} {\mnras}\ }\textbf {\bibinfo {volume} {497}},\ \bibinfo {pages}
  {3221} (\bibinfo {year} {2020})},\ \Eprint {https://arxiv.org/abs/2005.14208}
  {arXiv:2005.14208 [astro-ph.HE]} \BibitemShut {NoStop}%
\bibitem [{\citenamefont {{Metzger}}\ and\ \citenamefont
  {{Fern{\'a}ndez}}(2021)}]{metzger_2021_late-time}%
  \BibitemOpen
  \bibfield  {author} {\bibinfo {author} {\bibfnamefont {B.~D.}\ \bibnamefont
  {{Metzger}}}\ and\ \bibinfo {author} {\bibfnamefont {R.}~\bibnamefont
  {{Fern{\'a}ndez}}},\ }\bibfield  {title} {\bibinfo {title} {{From Neutrino-
  to Photon-cooled in Three Years: Can Fallback Accretion Explain the X-Ray
  Excess in GW170817?}},\ }\href {https://doi.org/10.3847/2041-8213/ac1169}
  {\bibfield  {journal} {\bibinfo  {journal} {\apj}\ }\textbf {\bibinfo
  {volume} {916}},\ \bibinfo {eid} {L3} (\bibinfo {year} {2021})},\ \Eprint
  {https://arxiv.org/abs/2106.02052} {arXiv:2106.02052 [astro-ph.HE]}
  \BibitemShut {NoStop}%
\bibitem [{\citenamefont {{Just}}\ \emph
  {et~al.}(2022{\natexlab{b}})\citenamefont {{Just}}, \citenamefont {{Abbar}},
  \citenamefont {{Wu}}, \citenamefont {{Tamborra}}, \citenamefont {{Janka}},\
  and\ \citenamefont {{Capozzi}}}]{Just2022_FFI}%
  \BibitemOpen
  \bibfield  {author} {\bibinfo {author} {\bibfnamefont {O.}~\bibnamefont
  {{Just}}}, \bibinfo {author} {\bibfnamefont {S.}~\bibnamefont {{Abbar}}},
  \bibinfo {author} {\bibfnamefont {M.-R.}\ \bibnamefont {{Wu}}}, \bibinfo
  {author} {\bibfnamefont {I.}~\bibnamefont {{Tamborra}}}, \bibinfo {author}
  {\bibfnamefont {H.-T.}\ \bibnamefont {{Janka}}},\ and\ \bibinfo {author}
  {\bibfnamefont {F.}~\bibnamefont {{Capozzi}}},\ }\bibfield  {title} {\bibinfo
  {title} {{Fast Neutrino Conversion in Hydrodynamic Simulations of
  Neutrino-Cooled Accretion Disks}},\ }\href@noop {} {\bibfield  {journal}
  {\bibinfo  {journal} {PRD, submitted}\ ,\ \bibinfo {eid} {arXiv:2203.16559}}
  (\bibinfo {year} {2022}{\natexlab{b}})},\ \Eprint
  {https://arxiv.org/abs/2203.16559} {arXiv:2203.16559 [astro-ph.HE]}
  \BibitemShut {NoStop}%
\bibitem [{\citenamefont {{Fujibayashi}}\ \emph {et~al.}(2022)\citenamefont
  {{Fujibayashi}}, \citenamefont {{Kiuchi}}, \citenamefont {{Wanajo}},
  \citenamefont {{Kyutoku}}, \citenamefont {{Sekiguchi}},\ and\ \citenamefont
  {{Shibata}}}]{fujibayashi_2022}%
  \BibitemOpen
  \bibfield  {author} {\bibinfo {author} {\bibfnamefont {S.}~\bibnamefont
  {{Fujibayashi}}}, \bibinfo {author} {\bibfnamefont {K.}~\bibnamefont
  {{Kiuchi}}}, \bibinfo {author} {\bibfnamefont {S.}~\bibnamefont {{Wanajo}}},
  \bibinfo {author} {\bibfnamefont {K.}~\bibnamefont {{Kyutoku}}}, \bibinfo
  {author} {\bibfnamefont {Y.}~\bibnamefont {{Sekiguchi}}},\ and\ \bibinfo
  {author} {\bibfnamefont {M.}~\bibnamefont {{Shibata}}},\ }\bibfield  {title}
  {\bibinfo {title} {{Comprehensive study of mass ejection and nucleosynthesis
  in binary neutron star mergers leaving short-lived massive neutron stars}},\
  }\href@noop {} {\bibfield  {journal} {\bibinfo  {journal} {arXiv e-prints}\
  ,\ \bibinfo {eid} {arXiv:2205.05557}} (\bibinfo {year} {2022})},\ \Eprint
  {https://arxiv.org/abs/2205.05557} {arXiv:2205.05557 [astro-ph.HE]}
  \BibitemShut {NoStop}%
\bibitem [{\citenamefont {{Fern{\'a}ndez}}\ \emph {et~al.}(2022)\citenamefont
  {{Fern{\'a}ndez}}, \citenamefont {{Richers}}, \citenamefont {{Mulyk}},\ and\
  \citenamefont {{Fahlman}}}]{fernandez_2022_FFI}%
  \BibitemOpen
  \bibfield  {author} {\bibinfo {author} {\bibfnamefont {R.}~\bibnamefont
  {{Fern{\'a}ndez}}}, \bibinfo {author} {\bibfnamefont {S.}~\bibnamefont
  {{Richers}}}, \bibinfo {author} {\bibfnamefont {N.}~\bibnamefont {{Mulyk}}},\
  and\ \bibinfo {author} {\bibfnamefont {S.}~\bibnamefont {{Fahlman}}},\
  }\bibfield  {title} {\bibinfo {title} {{Fast flavor instability in
  hypermassive neutron star disk outflows}},\ }\href
  {https://doi.org/10.1103/PhysRevD.106.103003} {\bibfield  {journal} {\bibinfo
   {journal} {\prd}\ }\textbf {\bibinfo {volume} {106}},\ \bibinfo {eid}
  {103003} (\bibinfo {year} {2022})},\ \Eprint
  {https://arxiv.org/abs/2207.10680} {arXiv:2207.10680 [astro-ph.HE]}
  \BibitemShut {NoStop}%
\bibitem [{\citenamefont {{Haddadi}}\ \emph {et~al.}(2022)\citenamefont
  {{Haddadi}}, \citenamefont {{Duez}}, \citenamefont {{Foucart}}, \citenamefont
  {{Ramirez}}, \citenamefont {{Fernandez}}, \citenamefont {{Knight}},
  \citenamefont {{Jesse}}, \citenamefont {{Hebert}}, \citenamefont {{Kidder}},
  \citenamefont {{Pfeiffer}},\ and\ \citenamefont {{Scheel}}}]{haddadi_2023}%
  \BibitemOpen
  \bibfield  {author} {\bibinfo {author} {\bibfnamefont {M.}~\bibnamefont
  {{Haddadi}}}, \bibinfo {author} {\bibfnamefont {M.~D.}\ \bibnamefont
  {{Duez}}}, \bibinfo {author} {\bibfnamefont {F.}~\bibnamefont {{Foucart}}},
  \bibinfo {author} {\bibfnamefont {T.}~\bibnamefont {{Ramirez}}}, \bibinfo
  {author} {\bibfnamefont {R.}~\bibnamefont {{Fernandez}}}, \bibinfo {author}
  {\bibfnamefont {A.~L.}\ \bibnamefont {{Knight}}}, \bibinfo {author}
  {\bibfnamefont {J.}~\bibnamefont {{Jesse}}}, \bibinfo {author} {\bibfnamefont
  {F.}~\bibnamefont {{Hebert}}}, \bibinfo {author} {\bibfnamefont {L.~E.}\
  \bibnamefont {{Kidder}}}, \bibinfo {author} {\bibfnamefont {H.~P.}\
  \bibnamefont {{Pfeiffer}}},\ and\ \bibinfo {author} {\bibfnamefont {M.~A.}\
  \bibnamefont {{Scheel}}},\ }\bibfield  {title} {\bibinfo {title} {{Late-time
  post-merger modeling of a compact binary: effects of relativity, r-process
  heating, and treatment of transport effects}},\ }\href@noop {} {\bibfield
  {journal} {\bibinfo  {journal} {CQG, submitted}\ ,\ \bibinfo {eid}
  {arXiv:2208.02367}} (\bibinfo {year} {2022})},\ \Eprint
  {https://arxiv.org/abs/2208.02367} {arXiv:2208.02367 [gr-qc]} \BibitemShut
  {NoStop}%
\bibitem [{\citenamefont {{Fern{\'a}ndez}}\ \emph {et~al.}(2019)\citenamefont
  {{Fern{\'a}ndez}}, \citenamefont {{Tchekhovskoy}}, \citenamefont
  {{Quataert}}, \citenamefont {{Foucart}},\ and\ \citenamefont
  {{Kasen}}}]{F19_grmhd}%
  \BibitemOpen
  \bibfield  {author} {\bibinfo {author} {\bibfnamefont {R.}~\bibnamefont
  {{Fern{\'a}ndez}}}, \bibinfo {author} {\bibfnamefont {A.}~\bibnamefont
  {{Tchekhovskoy}}}, \bibinfo {author} {\bibfnamefont {E.}~\bibnamefont
  {{Quataert}}}, \bibinfo {author} {\bibfnamefont {F.}~\bibnamefont
  {{Foucart}}},\ and\ \bibinfo {author} {\bibfnamefont {D.}~\bibnamefont
  {{Kasen}}},\ }\bibfield  {title} {\bibinfo {title} {{Long-term GRMHD
  simulations of neutron star merger accretion discs: implications for
  electromagnetic counterparts}},\ }\href
  {https://doi.org/10.1093/mnras/sty2932} {\bibfield  {journal} {\bibinfo
  {journal} {\mnras}\ }\textbf {\bibinfo {volume} {482}},\ \bibinfo {pages}
  {3373} (\bibinfo {year} {2019})},\ \Eprint {https://arxiv.org/abs/1808.00461}
  {arXiv:1808.00461 [astro-ph.HE]} \BibitemShut {NoStop}%
\bibitem [{\citenamefont {{Messer}}\ \emph {et~al.}(1998)\citenamefont
  {{Messer}}, \citenamefont {{Mezzacappa}}, \citenamefont {{Bruenn}},\ and\
  \citenamefont {{Guidry}}}]{messer_1998}%
  \BibitemOpen
  \bibfield  {author} {\bibinfo {author} {\bibfnamefont {O.~E.~B.}\
  \bibnamefont {{Messer}}}, \bibinfo {author} {\bibfnamefont {A.}~\bibnamefont
  {{Mezzacappa}}}, \bibinfo {author} {\bibfnamefont {S.~W.}\ \bibnamefont
  {{Bruenn}}},\ and\ \bibinfo {author} {\bibfnamefont {M.~W.}\ \bibnamefont
  {{Guidry}}},\ }\bibfield  {title} {\bibinfo {title} {{A Comparison of
  Boltzmann and Multigroup Flux-limited Diffusion Neutrino Transport during the
  Postbounce Shock Reheating Phase in Core-Collapse Supernovae}},\ }\href
  {https://doi.org/10.1086/306323} {\bibfield  {journal} {\bibinfo  {journal}
  {\apj}\ }\textbf {\bibinfo {volume} {507}},\ \bibinfo {pages} {353} (\bibinfo
  {year} {1998})}\BibitemShut {NoStop}%
\bibitem [{\citenamefont {{Yamada}}\ \emph {et~al.}(1999)\citenamefont
  {{Yamada}}, \citenamefont {{Janka}},\ and\ \citenamefont
  {{Suzuki}}}]{yamada_1999}%
  \BibitemOpen
  \bibfield  {author} {\bibinfo {author} {\bibfnamefont {S.}~\bibnamefont
  {{Yamada}}}, \bibinfo {author} {\bibfnamefont {H.-T.}\ \bibnamefont
  {{Janka}}},\ and\ \bibinfo {author} {\bibfnamefont {H.}~\bibnamefont
  {{Suzuki}}},\ }\bibfield  {title} {\bibinfo {title} {{Neutrino transport in
  type II supernovae: Boltzmann solver vs. Monte Carlo method}},\ }\href@noop
  {} {\bibfield  {journal} {\bibinfo  {journal} {A\&A}\ }\textbf {\bibinfo
  {volume} {344}},\ \bibinfo {pages} {533} (\bibinfo {year} {1999})},\ \Eprint
  {https://arxiv.org/abs/astro-ph/9809009} {arXiv:astro-ph/9809009 [astro-ph]}
  \BibitemShut {NoStop}%
\bibitem [{\citenamefont {{Liebend{\"o}rfer}}\ \emph
  {et~al.}(2005)\citenamefont {{Liebend{\"o}rfer}}, \citenamefont {{Rampp}},
  \citenamefont {{Janka}},\ and\ \citenamefont
  {{Mezzacappa}}}]{liebendoerfer_2005}%
  \BibitemOpen
  \bibfield  {author} {\bibinfo {author} {\bibfnamefont {M.}~\bibnamefont
  {{Liebend{\"o}rfer}}}, \bibinfo {author} {\bibfnamefont {M.}~\bibnamefont
  {{Rampp}}}, \bibinfo {author} {\bibfnamefont {H.~T.}\ \bibnamefont
  {{Janka}}},\ and\ \bibinfo {author} {\bibfnamefont {A.}~\bibnamefont
  {{Mezzacappa}}},\ }\bibfield  {title} {\bibinfo {title} {{Supernova
  Simulations with Boltzmann Neutrino Transport: A Comparison of Methods}},\
  }\href {https://doi.org/10.1086/427203} {\bibfield  {journal} {\bibinfo
  {journal} {\apj}\ }\textbf {\bibinfo {volume} {620}},\ \bibinfo {pages} {840}
  (\bibinfo {year} {2005})},\ \Eprint {https://arxiv.org/abs/astro-ph/0310662}
  {arXiv:astro-ph/0310662 [astro-ph]} \BibitemShut {NoStop}%
\bibitem [{\citenamefont {{Richers}}\ \emph {et~al.}(2017)\citenamefont
  {{Richers}}, \citenamefont {{Nagakura}}, \citenamefont {{Ott}}, \citenamefont
  {{Dolence}}, \citenamefont {{Sumiyoshi}},\ and\ \citenamefont
  {{Yamada}}}]{richers_2017}%
  \BibitemOpen
  \bibfield  {author} {\bibinfo {author} {\bibfnamefont {S.}~\bibnamefont
  {{Richers}}}, \bibinfo {author} {\bibfnamefont {H.}~\bibnamefont
  {{Nagakura}}}, \bibinfo {author} {\bibfnamefont {C.~D.}\ \bibnamefont
  {{Ott}}}, \bibinfo {author} {\bibfnamefont {J.}~\bibnamefont {{Dolence}}},
  \bibinfo {author} {\bibfnamefont {K.}~\bibnamefont {{Sumiyoshi}}},\ and\
  \bibinfo {author} {\bibfnamefont {S.}~\bibnamefont {{Yamada}}},\ }\bibfield
  {title} {\bibinfo {title} {{A Detailed Comparison of Multidimensional
  Boltzmann Neutrino Transport Methods in Core-collapse Supernovae}},\ }\href
  {https://doi.org/10.3847/1538-4357/aa8bb2} {\bibfield  {journal} {\bibinfo
  {journal} {\apj}\ }\textbf {\bibinfo {volume} {847}},\ \bibinfo {eid} {133}
  (\bibinfo {year} {2017})},\ \Eprint {https://arxiv.org/abs/1706.06187}
  {arXiv:1706.06187 [astro-ph.HE]} \BibitemShut {NoStop}%
\bibitem [{\citenamefont {{Just}}\ \emph {et~al.}(2018)\citenamefont {{Just}},
  \citenamefont {{Bollig}}, \citenamefont {{Janka}}, \citenamefont
  {{Obergaulinger}}, \citenamefont {{Glas}},\ and\ \citenamefont
  {{Nagataki}}}]{just_2018}%
  \BibitemOpen
  \bibfield  {author} {\bibinfo {author} {\bibfnamefont {O.}~\bibnamefont
  {{Just}}}, \bibinfo {author} {\bibfnamefont {R.}~\bibnamefont {{Bollig}}},
  \bibinfo {author} {\bibfnamefont {H.~T.}\ \bibnamefont {{Janka}}}, \bibinfo
  {author} {\bibfnamefont {M.}~\bibnamefont {{Obergaulinger}}}, \bibinfo
  {author} {\bibfnamefont {R.}~\bibnamefont {{Glas}}},\ and\ \bibinfo {author}
  {\bibfnamefont {S.}~\bibnamefont {{Nagataki}}},\ }\bibfield  {title}
  {\bibinfo {title} {{Core-collapse supernova simulations in one and two
  dimensions: comparison of codes and approximations}},\ }\href
  {https://doi.org/10.1093/mnras/sty2578} {\bibfield  {journal} {\bibinfo
  {journal} {\mnras}\ }\textbf {\bibinfo {volume} {481}},\ \bibinfo {pages}
  {4786} (\bibinfo {year} {2018})},\ \Eprint {https://arxiv.org/abs/1805.03953}
  {arXiv:1805.03953 [astro-ph.HE]} \BibitemShut {NoStop}%
\bibitem [{\citenamefont {{O'Connor}}\ \emph {et~al.}(2018)\citenamefont
  {{O'Connor}}, \citenamefont {{Bollig}}, \citenamefont {{Burrows}},
  \citenamefont {{Couch}}, \citenamefont {{Fischer}}, \citenamefont {{Janka}},
  \citenamefont {{Kotake}}, \citenamefont {{Lentz}}, \citenamefont
  {{Liebend{\"o}rfer}}, \citenamefont {{Messer}}, \citenamefont {{Mezzacappa}},
  \citenamefont {{Takiwaki}},\ and\ \citenamefont
  {{Vartanyan}}}]{oconnor_2018}%
  \BibitemOpen
  \bibfield  {author} {\bibinfo {author} {\bibfnamefont {E.}~\bibnamefont
  {{O'Connor}}}, \bibinfo {author} {\bibfnamefont {R.}~\bibnamefont
  {{Bollig}}}, \bibinfo {author} {\bibfnamefont {A.}~\bibnamefont {{Burrows}}},
  \bibinfo {author} {\bibfnamefont {S.}~\bibnamefont {{Couch}}}, \bibinfo
  {author} {\bibfnamefont {T.}~\bibnamefont {{Fischer}}}, \bibinfo {author}
  {\bibfnamefont {H.-T.}\ \bibnamefont {{Janka}}}, \bibinfo {author}
  {\bibfnamefont {K.}~\bibnamefont {{Kotake}}}, \bibinfo {author}
  {\bibfnamefont {E.~J.}\ \bibnamefont {{Lentz}}}, \bibinfo {author}
  {\bibfnamefont {M.}~\bibnamefont {{Liebend{\"o}rfer}}}, \bibinfo {author}
  {\bibfnamefont {O.~E.~B.}\ \bibnamefont {{Messer}}}, \bibinfo {author}
  {\bibfnamefont {A.}~\bibnamefont {{Mezzacappa}}}, \bibinfo {author}
  {\bibfnamefont {T.}~\bibnamefont {{Takiwaki}}},\ and\ \bibinfo {author}
  {\bibfnamefont {D.}~\bibnamefont {{Vartanyan}}},\ }\bibfield  {title}
  {\bibinfo {title} {{Global comparison of core-collapse supernova simulations
  in spherical symmetry}},\ }\href {https://doi.org/10.1088/1361-6471/aadeae}
  {\bibfield  {journal} {\bibinfo  {journal} {Journal of Physics G Nuclear
  Physics}\ }\textbf {\bibinfo {volume} {45}},\ \bibinfo {pages} {104001}
  (\bibinfo {year} {2018})},\ \Eprint {https://arxiv.org/abs/1806.04175}
  {arXiv:1806.04175 [astro-ph.HE]} \BibitemShut {NoStop}%
\bibitem [{\citenamefont {{Cabez{\'o}n}}\ \emph {et~al.}(2018)\citenamefont
  {{Cabez{\'o}n}}, \citenamefont {{Pan}}, \citenamefont {{Liebend{\"o}rfer}},
  \citenamefont {{Kuroda}}, \citenamefont {{Ebinger}}, \citenamefont
  {{Heinimann}}, \citenamefont {{Perego}},\ and\ \citenamefont
  {{Thielemann}}}]{cabezon_2018}%
  \BibitemOpen
  \bibfield  {author} {\bibinfo {author} {\bibfnamefont {R.~M.}\ \bibnamefont
  {{Cabez{\'o}n}}}, \bibinfo {author} {\bibfnamefont {K.-C.}\ \bibnamefont
  {{Pan}}}, \bibinfo {author} {\bibfnamefont {M.}~\bibnamefont
  {{Liebend{\"o}rfer}}}, \bibinfo {author} {\bibfnamefont {T.}~\bibnamefont
  {{Kuroda}}}, \bibinfo {author} {\bibfnamefont {K.}~\bibnamefont {{Ebinger}}},
  \bibinfo {author} {\bibfnamefont {O.}~\bibnamefont {{Heinimann}}}, \bibinfo
  {author} {\bibfnamefont {A.}~\bibnamefont {{Perego}}},\ and\ \bibinfo
  {author} {\bibfnamefont {F.-K.}\ \bibnamefont {{Thielemann}}},\ }\bibfield
  {title} {\bibinfo {title} {{Core-collapse supernovae in the hall of mirrors.
  A three-dimensional code-comparison project}},\ }\href
  {https://doi.org/10.1051/0004-6361/201833705} {\bibfield  {journal} {\bibinfo
   {journal} {A\&A}\ }\textbf {\bibinfo {volume} {619}},\ \bibinfo {eid} {A118}
  (\bibinfo {year} {2018})},\ \Eprint {https://arxiv.org/abs/1806.09184}
  {arXiv:1806.09184 [astro-ph.HE]} \BibitemShut {NoStop}%
\bibitem [{\citenamefont {{Varma}}\ \emph {et~al.}(2021)\citenamefont
  {{Varma}}, \citenamefont {{M{\"u}ller}},\ and\ \citenamefont
  {{Obergaulinger}}}]{varma_2021}%
  \BibitemOpen
  \bibfield  {author} {\bibinfo {author} {\bibfnamefont {V.}~\bibnamefont
  {{Varma}}}, \bibinfo {author} {\bibfnamefont {B.}~\bibnamefont
  {{M{\"u}ller}}},\ and\ \bibinfo {author} {\bibfnamefont {M.}~\bibnamefont
  {{Obergaulinger}}},\ }\bibfield  {title} {\bibinfo {title} {{A comparison of
  2D Magnetohydrodynamic supernova simulations with the COCONUT-FMT and
  AENUS-ALCAR codes}},\ }\href {https://doi.org/10.1093/mnras/stab2983}
  {\bibfield  {journal} {\bibinfo  {journal} {\mnras}\ }\textbf {\bibinfo
  {volume} {508}},\ \bibinfo {pages} {6033} (\bibinfo {year} {2021})},\ \Eprint
  {https://arxiv.org/abs/2109.03603} {arXiv:2109.03603 [astro-ph.HE]}
  \BibitemShut {NoStop}%
\bibitem [{\citenamefont {{Just}}\ \emph
  {et~al.}(2015{\natexlab{b}})\citenamefont {{Just}}, \citenamefont
  {{Obergaulinger}},\ and\ \citenamefont {{Janka}}}]{Just2015b}%
  \BibitemOpen
  \bibfield  {author} {\bibinfo {author} {\bibfnamefont {O.}~\bibnamefont
  {{Just}}}, \bibinfo {author} {\bibfnamefont {M.}~\bibnamefont
  {{Obergaulinger}}},\ and\ \bibinfo {author} {\bibfnamefont {H.-T.}\
  \bibnamefont {{Janka}}},\ }\bibfield  {title} {\bibinfo {title} {{A new
  multidimensional, energy-dependent two-moment transport code for
  neutrino-hydrodynamics}},\ }\href@noop {} {\bibfield  {journal} {\bibinfo
  {journal} {\mnras}\ }\textbf {\bibinfo {volume} {453}},\ \bibinfo {pages}
  {3386} (\bibinfo {year} {2015}{\natexlab{b}})}\BibitemShut {NoStop}%
\bibitem [{\citenamefont {Obergaulinger}(2008)}]{Obergaulinger2008a}%
  \BibitemOpen
  \bibfield  {author} {\bibinfo {author} {\bibfnamefont {M.}~\bibnamefont
  {Obergaulinger}},\ }\emph {\bibinfo {title} {Astrophysical
  magnetohydrodynamics and radiative transfer}},\ \href@noop {} {\bibinfo
  {type} {Dissertation}},\ \bibinfo  {school} {Technische Universit{\"a}t
  M{\"u}nchen}, \bibinfo {address} {M{\"u}nchen} (\bibinfo {year}
  {2008})\BibitemShut {NoStop}%
\bibitem [{\citenamefont {{Mignone}}(2014)}]{Mignone2014a}%
  \BibitemOpen
  \bibfield  {author} {\bibinfo {author} {\bibfnamefont {A.}~\bibnamefont
  {{Mignone}}},\ }\bibfield  {title} {\bibinfo {title} {{High-order
  conservative reconstruction schemes for finite volume methods in cylindrical
  and spherical coordinates}},\ }\href@noop {} {\bibfield  {journal} {\bibinfo
  {journal} {Journal of Computational Physics}\ }\textbf {\bibinfo {volume}
  {270}},\ \bibinfo {pages} {784} (\bibinfo {year} {2014})}\BibitemShut
  {NoStop}%
\bibitem [{\citenamefont {{Artemova}}\ \emph {et~al.}(1996)\citenamefont
  {{Artemova}}, \citenamefont {{Bjoernsson}},\ and\ \citenamefont
  {{Novikov}}}]{artemova1996}%
  \BibitemOpen
  \bibfield  {author} {\bibinfo {author} {\bibfnamefont {I.~V.}\ \bibnamefont
  {{Artemova}}}, \bibinfo {author} {\bibfnamefont {G.}~\bibnamefont
  {{Bjoernsson}}},\ and\ \bibinfo {author} {\bibfnamefont {I.~D.}\ \bibnamefont
  {{Novikov}}},\ }\bibfield  {title} {\bibinfo {title} {{Modified Newtonian
  Potentials for the Description of Relativistic Effects in Accretion Disks
  around Black Holes}},\ }\href {https://doi.org/10.1086/177084} {\bibfield
  {journal} {\bibinfo  {journal} {ApJ}\ }\textbf {\bibinfo {volume} {461}},\
  \bibinfo {pages} {565} (\bibinfo {year} {1996})}\BibitemShut {NoStop}%
\bibitem [{\citenamefont {{Janka}}\ and\ \citenamefont
  {{Mueller}}(1996)}]{Janka1996b}%
  \BibitemOpen
  \bibfield  {author} {\bibinfo {author} {\bibfnamefont {H.~T.}\ \bibnamefont
  {{Janka}}}\ and\ \bibinfo {author} {\bibfnamefont {E.}~\bibnamefont
  {{Mueller}}},\ }\bibfield  {title} {\bibinfo {title} {{Neutrino heating,
  convection, and the mechanism of Type-II supernova explosions.}},\
  }\href@noop {} {\bibfield  {journal} {\bibinfo  {journal} {\aap}\ }\textbf
  {\bibinfo {volume} {306}},\ \bibinfo {pages} {167} (\bibinfo {year}
  {1996})}\BibitemShut {NoStop}%
\bibitem [{\citenamefont {{Minerbo}}(1978)}]{Minerbo1978}%
  \BibitemOpen
  \bibfield  {author} {\bibinfo {author} {\bibfnamefont {G.~N.}\ \bibnamefont
  {{Minerbo}}},\ }\bibfield  {title} {\bibinfo {title} {{Maximum entropy
  Eddington factors.}},\ }\href@noop {} {\bibfield  {journal} {\bibinfo
  {journal} {Journal of Quantitative Spectroscopy and Radiative Transfer}\
  }\textbf {\bibinfo {volume} {20}},\ \bibinfo {pages} {541} (\bibinfo {year}
  {1978})}\BibitemShut {NoStop}%
\bibitem [{\citenamefont {{Bruenn}}(1985)}]{Bruenn1985}%
  \BibitemOpen
  \bibfield  {author} {\bibinfo {author} {\bibfnamefont {S.~W.}\ \bibnamefont
  {{Bruenn}}},\ }\bibfield  {title} {\bibinfo {title} {{Stellar core collapse -
  Numerical model and infall epoch}},\ }\href@noop {} {\bibfield  {journal}
  {\bibinfo  {journal} {Astrophysical Journal Supplement}\ }\textbf {\bibinfo
  {volume} {58}},\ \bibinfo {pages} {771} (\bibinfo {year} {1985})}\BibitemShut
  {NoStop}%
\bibitem [{\citenamefont {{Horowitz}}(2002)}]{Horowitz2002a}%
  \BibitemOpen
  \bibfield  {author} {\bibinfo {author} {\bibfnamefont {C.~J.}\ \bibnamefont
  {{Horowitz}}},\ }\bibfield  {title} {\bibinfo {title} {{Weak magnetism for
  antineutrinos in supernovae}},\ }\href@noop {} {\bibfield  {journal}
  {\bibinfo  {journal} {\prd}\ }\textbf {\bibinfo {volume} {65}},\ \bibinfo
  {pages} {043001} (\bibinfo {year} {2002})}\BibitemShut {NoStop}%
\bibitem [{\citenamefont {{Pons}}\ \emph {et~al.}(1998)\citenamefont {{Pons}},
  \citenamefont {{Miralles}},\ and\ \citenamefont {{Ibanez}}}]{Pons1998}%
  \BibitemOpen
  \bibfield  {author} {\bibinfo {author} {\bibfnamefont {J.~A.}\ \bibnamefont
  {{Pons}}}, \bibinfo {author} {\bibfnamefont {J.~A.}\ \bibnamefont
  {{Miralles}}},\ and\ \bibinfo {author} {\bibfnamefont {J.~M.~A.}\
  \bibnamefont {{Ibanez}}},\ }\bibfield  {title} {\bibinfo {title} {{Legendre
  expansion of the $\nu\bar\nu$-annihilation}},\ }\href@noop {} {\bibfield
  {journal} {\bibinfo  {journal} {Astronomy \& Astrophysics Supplement}\
  }\textbf {\bibinfo {volume} {129}},\ \bibinfo {pages} {343} (\bibinfo {year}
  {1998})}\BibitemShut {NoStop}%
\bibitem [{\citenamefont {{Hannestad}}\ and\ \citenamefont
  {{Raffelt}}(1998)}]{Hannestad1998}%
  \BibitemOpen
  \bibfield  {author} {\bibinfo {author} {\bibfnamefont {S.}~\bibnamefont
  {{Hannestad}}}\ and\ \bibinfo {author} {\bibfnamefont {G.}~\bibnamefont
  {{Raffelt}}},\ }\bibfield  {title} {\bibinfo {title} {{Supernova Neutrino
  Opacity from Nucleon-Nucleon Bremsstrahlung and Related Processes}},\
  }\href@noop {} {\bibfield  {journal} {\bibinfo  {journal} {\apj}\ }\textbf
  {\bibinfo {volume} {507}},\ \bibinfo {pages} {339} (\bibinfo {year}
  {1998})}\BibitemShut {NoStop}%
\bibitem [{\citenamefont {O'Connor}(2015)}]{OConnor2015a}%
  \BibitemOpen
  \bibfield  {author} {\bibinfo {author} {\bibfnamefont {E.}~\bibnamefont
  {O'Connor}},\ }\bibfield  {title} {\bibinfo {title} {An open-source neutrino
  radiation hydrodynamics code for core-collapse supernovae},\ }\href@noop {}
  {\bibfield  {journal} {\bibinfo  {journal} {Astrophysical Journal
  Supplement}\ }\textbf {\bibinfo {volume} {219}},\ \bibinfo {pages} {24}
  (\bibinfo {year} {2015})}\BibitemShut {NoStop}%
\bibitem [{\citenamefont {{Just}}\ \emph {et~al.}(2016)\citenamefont {{Just}},
  \citenamefont {{Obergaulinger}}, \citenamefont {{Janka}}, \citenamefont
  {{Bauswein}},\ and\ \citenamefont {{Schwarz}}}]{Just2016}%
  \BibitemOpen
  \bibfield  {author} {\bibinfo {author} {\bibfnamefont {O.}~\bibnamefont
  {{Just}}}, \bibinfo {author} {\bibfnamefont {M.}~\bibnamefont
  {{Obergaulinger}}}, \bibinfo {author} {\bibfnamefont {H.-T.}\ \bibnamefont
  {{Janka}}}, \bibinfo {author} {\bibfnamefont {A.}~\bibnamefont
  {{Bauswein}}},\ and\ \bibinfo {author} {\bibfnamefont {N.}~\bibnamefont
  {{Schwarz}}},\ }\bibfield  {title} {\bibinfo {title} {{Neutron-star Merger
  Ejecta as Obstacles to Neutrino-powered Jets of Gamma-Ray Bursts}},\
  }\href@noop {} {\bibfield  {journal} {\bibinfo  {journal} {\apj}\ }\textbf
  {\bibinfo {volume} {816}},\ \bibinfo {pages} {L30} (\bibinfo {year}
  {2016})}\BibitemShut {NoStop}%
\bibitem [{\citenamefont {{Fryxell}}\ \emph {et~al.}(2000)\citenamefont
  {{Fryxell}}, \citenamefont {{Olson}}, \citenamefont {{Ricker}}, \citenamefont
  {{Timmes}}, \citenamefont {{Zingale}}, \citenamefont {{Lamb}}, \citenamefont
  {{MacNeice}}, \citenamefont {{Rosner}}, \citenamefont {{Truran}},\ and\
  \citenamefont {{Tufo}}}]{fryxell00}%
  \BibitemOpen
  \bibfield  {author} {\bibinfo {author} {\bibfnamefont {B.}~\bibnamefont
  {{Fryxell}}}, \bibinfo {author} {\bibfnamefont {K.}~\bibnamefont {{Olson}}},
  \bibinfo {author} {\bibfnamefont {P.}~\bibnamefont {{Ricker}}}, \bibinfo
  {author} {\bibfnamefont {F.~X.}\ \bibnamefont {{Timmes}}}, \bibinfo {author}
  {\bibfnamefont {M.}~\bibnamefont {{Zingale}}}, \bibinfo {author}
  {\bibfnamefont {D.~Q.}\ \bibnamefont {{Lamb}}}, \bibinfo {author}
  {\bibfnamefont {P.}~\bibnamefont {{MacNeice}}}, \bibinfo {author}
  {\bibfnamefont {R.}~\bibnamefont {{Rosner}}}, \bibinfo {author}
  {\bibfnamefont {J.~W.}\ \bibnamefont {{Truran}}},\ and\ \bibinfo {author}
  {\bibfnamefont {H.}~\bibnamefont {{Tufo}}},\ }\bibfield  {title} {\bibinfo
  {title} {{FLASH: An Adaptive Mesh Hydrodynamics Code for Modeling
  Astrophysical Thermonuclear Flashes}},\ }\href
  {https://doi.org/10.1086/317361} {\bibfield  {journal} {\bibinfo  {journal}
  {ApJS}\ }\textbf {\bibinfo {volume} {131}},\ \bibinfo {pages} {273} (\bibinfo
  {year} {2000})}\BibitemShut {NoStop}%
\bibitem [{\citenamefont {Dubey}\ \emph {et~al.}(2009)\citenamefont {Dubey},
  \citenamefont {Antypas}, \citenamefont {Ganapathy}, \citenamefont {Reid},
  \citenamefont {Riley}, \citenamefont {Sheeler}, \citenamefont {Siegel},\ and\
  \citenamefont {Weide}}]{dubey2009}%
  \BibitemOpen
  \bibfield  {author} {\bibinfo {author} {\bibfnamefont {A.}~\bibnamefont
  {Dubey}}, \bibinfo {author} {\bibfnamefont {K.}~\bibnamefont {Antypas}},
  \bibinfo {author} {\bibfnamefont {M.~K.}\ \bibnamefont {Ganapathy}}, \bibinfo
  {author} {\bibfnamefont {L.~B.}\ \bibnamefont {Reid}}, \bibinfo {author}
  {\bibfnamefont {K.}~\bibnamefont {Riley}}, \bibinfo {author} {\bibfnamefont
  {D.}~\bibnamefont {Sheeler}}, \bibinfo {author} {\bibfnamefont
  {A.}~\bibnamefont {Siegel}},\ and\ \bibinfo {author} {\bibfnamefont
  {K.}~\bibnamefont {Weide}},\ }\bibfield  {title} {\bibinfo {title}
  {Extensible component-based architecture for flash, a massively parallel,
  multiphysics simulation code},\ }\href {https://doi.org/DOI:
  10.1016/j.parco.2009.08.001} {\bibfield  {journal} {\bibinfo  {journal} {J.
  Par. Comp.}\ }\textbf {\bibinfo {volume} {35}},\ \bibinfo {pages} {512 }
  (\bibinfo {year} {2009})}\BibitemShut {NoStop}%
\bibitem [{\citenamefont {{Colella}}\ and\ \citenamefont
  {{Woodward}}(1984)}]{colella84}%
  \BibitemOpen
  \bibfield  {author} {\bibinfo {author} {\bibfnamefont {P.}~\bibnamefont
  {{Colella}}}\ and\ \bibinfo {author} {\bibfnamefont {P.~R.}\ \bibnamefont
  {{Woodward}}},\ }\bibfield  {title} {\bibinfo {title} {{The Piecewise
  Parabolic Method (PPM) for Gas-Dynamical Simulations}},\ }\href@noop {}
  {\bibfield  {journal} {\bibinfo  {journal} {JCP}\ }\textbf {\bibinfo {volume}
  {54}},\ \bibinfo {pages} {174} (\bibinfo {year} {1984})}\BibitemShut
  {NoStop}%
\bibitem [{\citenamefont {{Fern{\'a}ndez}}(2012)}]{F12}%
  \BibitemOpen
  \bibfield  {author} {\bibinfo {author} {\bibfnamefont {R.}~\bibnamefont
  {{Fern{\'a}ndez}}},\ }\bibfield  {title} {\bibinfo {title} {{Hydrodynamics of
  Core-collapse Supernovae at the Transition to Explosion. I. Spherical
  Symmetry}},\ }\href {https://doi.org/10.1088/0004-637X/749/2/142} {\bibfield
  {journal} {\bibinfo  {journal} {ApJ}\ }\textbf {\bibinfo {volume} {749}},\
  \bibinfo {eid} {142} (\bibinfo {year} {2012})},\ \Eprint
  {https://arxiv.org/abs/1111.0665} {arXiv:1111.0665 [astro-ph.HE]}
  \BibitemShut {NoStop}%
\bibitem [{\citenamefont {{Fern{\'a}ndez}}\ and\ \citenamefont
  {{Metzger}}(2013{\natexlab{b}})}]{FM12}%
  \BibitemOpen
  \bibfield  {author} {\bibinfo {author} {\bibfnamefont {R.}~\bibnamefont
  {{Fern{\'a}ndez}}}\ and\ \bibinfo {author} {\bibfnamefont {B.~D.}\
  \bibnamefont {{Metzger}}},\ }\bibfield  {title} {\bibinfo {title} {{Nuclear
  Dominated Accretion Flows in Two Dimensions. I. Torus Evolution with
  Parametric Microphysics}},\ }\href
  {https://doi.org/10.1088/0004-637X/763/2/108} {\bibfield  {journal} {\bibinfo
   {journal} {ApJ}\ }\textbf {\bibinfo {volume} {763}},\ \bibinfo {eid} {108}
  (\bibinfo {year} {2013}{\natexlab{b}})},\ \Eprint
  {https://arxiv.org/abs/1209.2712} {arXiv:1209.2712 [astro-ph.HE]}
  \BibitemShut {NoStop}%
\bibitem [{\citenamefont {{Lippuner}}\ \emph {et~al.}(2017)\citenamefont
  {{Lippuner}}, \citenamefont {{Fern{\'a}ndez}}, \citenamefont {{Roberts}},
  \citenamefont {{Foucart}}, \citenamefont {{Kasen}}, \citenamefont
  {{Metzger}},\ and\ \citenamefont {{Ott}}}]{lippuner_2017}%
  \BibitemOpen
  \bibfield  {author} {\bibinfo {author} {\bibfnamefont {J.}~\bibnamefont
  {{Lippuner}}}, \bibinfo {author} {\bibfnamefont {R.}~\bibnamefont
  {{Fern{\'a}ndez}}}, \bibinfo {author} {\bibfnamefont {L.~F.}\ \bibnamefont
  {{Roberts}}}, \bibinfo {author} {\bibfnamefont {F.}~\bibnamefont
  {{Foucart}}}, \bibinfo {author} {\bibfnamefont {D.}~\bibnamefont {{Kasen}}},
  \bibinfo {author} {\bibfnamefont {B.~D.}\ \bibnamefont {{Metzger}}},\ and\
  \bibinfo {author} {\bibfnamefont {C.~D.}\ \bibnamefont {{Ott}}},\ }\bibfield
  {title} {\bibinfo {title} {{Signatures of hypermassive neutron star lifetimes
  on r-process nucleosynthesis in the disc ejecta from neutron star mergers}},\
  }\href {https://doi.org/10.1093/mnras/stx1987} {\bibfield  {journal}
  {\bibinfo  {journal} {MNRAS}\ }\textbf {\bibinfo {volume} {472}},\ \bibinfo
  {pages} {904} (\bibinfo {year} {2017})},\ \Eprint
  {https://arxiv.org/abs/1703.06216} {arXiv:1703.06216 [astro-ph.HE]}
  \BibitemShut {NoStop}%
\bibitem [{\citenamefont {{Ruffert}}\ \emph {et~al.}(1996)\citenamefont
  {{Ruffert}}, \citenamefont {{Janka}},\ and\ \citenamefont
  {{Schaefer}}}]{ruffert_1996}%
  \BibitemOpen
  \bibfield  {author} {\bibinfo {author} {\bibfnamefont {M.}~\bibnamefont
  {{Ruffert}}}, \bibinfo {author} {\bibfnamefont {H.-T.}\ \bibnamefont
  {{Janka}}},\ and\ \bibinfo {author} {\bibfnamefont {G.}~\bibnamefont
  {{Schaefer}}},\ }\bibfield  {title} {\bibinfo {title} {{Coalescing neutron
  stars - a step towards physical models. I. Hydrodynamic evolution and
  gravitational-wave emission.}},\ }\href@noop {} {\bibfield  {journal}
  {\bibinfo  {journal} {A\&A}\ }\textbf {\bibinfo {volume} {311}},\ \bibinfo
  {pages} {532} (\bibinfo {year} {1996})},\ \Eprint
  {https://arxiv.org/abs/arXiv:astro-ph/9509006} {arXiv:astro-ph/9509006}
  \BibitemShut {NoStop}%
\bibitem [{\citenamefont {{Timmes}}\ and\ \citenamefont
  {{Swesty}}(2000)}]{timmes2000}%
  \BibitemOpen
  \bibfield  {author} {\bibinfo {author} {\bibfnamefont {F.~X.}\ \bibnamefont
  {{Timmes}}}\ and\ \bibinfo {author} {\bibfnamefont {F.~D.}\ \bibnamefont
  {{Swesty}}},\ }\bibfield  {title} {\bibinfo {title} {{The Accuracy,
  Consistency, and Speed of an Electron-Positron Equation of State Based on
  Table Interpolation of the Helmholtz Free Energy}},\ }\href
  {https://doi.org/10.1086/313304} {\bibfield  {journal} {\bibinfo  {journal}
  {ApJS}\ }\textbf {\bibinfo {volume} {126}},\ \bibinfo {pages} {501} (\bibinfo
  {year} {2000})}\BibitemShut {NoStop}%
\bibitem [{\citenamefont {Mendoza-Temis}\ \emph {et~al.}(2015)\citenamefont
  {Mendoza-Temis}, \citenamefont {Wu}, \citenamefont {Langanke}, \citenamefont
  {Mart\'{\i}nez-Pinedo}, \citenamefont {Bauswein},\ and\ \citenamefont
  {Janka}}]{Mendoza-Temis.Wu.ea:2015}%
  \BibitemOpen
  \bibfield  {author} {\bibinfo {author} {\bibfnamefont {J.~J.}\ \bibnamefont
  {Mendoza-Temis}}, \bibinfo {author} {\bibfnamefont {M.-R.}\ \bibnamefont
  {Wu}}, \bibinfo {author} {\bibfnamefont {K.}~\bibnamefont {Langanke}},
  \bibinfo {author} {\bibfnamefont {G.}~\bibnamefont {Mart\'{\i}nez-Pinedo}},
  \bibinfo {author} {\bibfnamefont {A.}~\bibnamefont {Bauswein}},\ and\
  \bibinfo {author} {\bibfnamefont {H.-T.}\ \bibnamefont {Janka}},\ }\bibfield
  {title} {\bibinfo {title} {Nuclear robustness of the $r$ process in
  neutron-star mergers},\ }\href {https://doi.org/10.1103/PhysRevC.92.055805}
  {\bibfield  {journal} {\bibinfo  {journal} {Phys. Rev. C}\ }\textbf {\bibinfo
  {volume} {92}},\ \bibinfo {pages} {055805} (\bibinfo {year}
  {2015})}\BibitemShut {NoStop}%
\bibitem [{\citenamefont {Just}\ \emph {et~al.}(2022)\citenamefont {Just},
  \citenamefont {Kullmann}, \citenamefont {Goriely}, \citenamefont {Bauswein},
  \citenamefont {Janka},\ and\ \citenamefont {Collins}}]{Just2022a}%
  \BibitemOpen
  \bibfield  {author} {\bibinfo {author} {\bibfnamefont {O.}~\bibnamefont
  {Just}}, \bibinfo {author} {\bibfnamefont {I.}~\bibnamefont {Kullmann}},
  \bibinfo {author} {\bibfnamefont {S.}~\bibnamefont {Goriely}}, \bibinfo
  {author} {\bibfnamefont {A.}~\bibnamefont {Bauswein}}, \bibinfo {author}
  {\bibfnamefont {H.-T.}\ \bibnamefont {Janka}},\ and\ \bibinfo {author}
  {\bibfnamefont {C.~E.}\ \bibnamefont {Collins}},\ }\bibfield  {title}
  {\bibinfo {title} {{Dynamical ejecta of neutron star mergers with nucleonic
  weak processes -- II: kilonova emission}},\ }\href@noop {} {\bibfield
  {journal} {\bibinfo  {journal} {\mnras}\ }\textbf {\bibinfo {volume} {510}},\
  \bibinfo {pages} {2820} (\bibinfo {year} {2022})}\BibitemShut {NoStop}%
\bibitem [{\citenamefont {{Barnes}}\ \emph {et~al.}(2016)\citenamefont
  {{Barnes}}, \citenamefont {{Kasen}}, \citenamefont {{Wu}},\ and\
  \citenamefont {{Mart{\'\i}nez-Pinedo}}}]{Barnes2016a}%
  \BibitemOpen
  \bibfield  {author} {\bibinfo {author} {\bibfnamefont {J.}~\bibnamefont
  {{Barnes}}}, \bibinfo {author} {\bibfnamefont {D.}~\bibnamefont {{Kasen}}},
  \bibinfo {author} {\bibfnamefont {M.-R.}\ \bibnamefont {{Wu}}},\ and\
  \bibinfo {author} {\bibfnamefont {G.}~\bibnamefont
  {{Mart{\'\i}nez-Pinedo}}},\ }\bibfield  {title} {\bibinfo {title}
  {{Radioactivity and Thermalization in the Ejecta of Compact Object Mergers
  and Their Impact on Kilonova Light Curves}},\ }\href@noop {} {\bibfield
  {journal} {\bibinfo  {journal} {\apj}\ }\textbf {\bibinfo {volume} {829}},\
  \bibinfo {pages} {110} (\bibinfo {year} {2016})}\BibitemShut {NoStop}%
\bibitem [{\citenamefont {{Wu}}\ \emph {et~al.}(2019)\citenamefont {{Wu}},
  \citenamefont {{Barnes}}, \citenamefont {{Mart{\'\i}nez-Pinedo}},\ and\
  \citenamefont {{Metzger}}}]{Wu2019c}%
  \BibitemOpen
  \bibfield  {author} {\bibinfo {author} {\bibfnamefont {M.-R.}\ \bibnamefont
  {{Wu}}}, \bibinfo {author} {\bibfnamefont {J.}~\bibnamefont {{Barnes}}},
  \bibinfo {author} {\bibfnamefont {G.}~\bibnamefont
  {{Mart{\'\i}nez-Pinedo}}},\ and\ \bibinfo {author} {\bibfnamefont {B.~D.}\
  \bibnamefont {{Metzger}}},\ }\bibfield  {title} {\bibinfo {title}
  {{Fingerprints of Heavy-Element Nucleosynthesis in the Late-Time Lightcurves
  of Kilonovae}},\ }\href@noop {} {\bibfield  {journal} {\bibinfo  {journal}
  {\prl}\ }\textbf {\bibinfo {volume} {122}},\ \bibinfo {eid} {062701}
  (\bibinfo {year} {2019})},\ \Eprint {https://arxiv.org/abs/1808.10459}
  {arXiv:1808.10459 [astro-ph.HE]} \BibitemShut {NoStop}%
\bibitem [{\citenamefont {{Most}}\ \emph {et~al.}(2021)\citenamefont {{Most}},
  \citenamefont {{Papenfort}}, \citenamefont {{Tootle}},\ and\ \citenamefont
  {{Rezzolla}}}]{most_2021_disk}%
  \BibitemOpen
  \bibfield  {author} {\bibinfo {author} {\bibfnamefont {E.~R.}\ \bibnamefont
  {{Most}}}, \bibinfo {author} {\bibfnamefont {L.~J.}\ \bibnamefont
  {{Papenfort}}}, \bibinfo {author} {\bibfnamefont {S.~D.}\ \bibnamefont
  {{Tootle}}},\ and\ \bibinfo {author} {\bibfnamefont {L.}~\bibnamefont
  {{Rezzolla}}},\ }\bibfield  {title} {\bibinfo {title} {{On accretion discs
  formed in MHD simulations of black hole-neutron star mergers with accurate
  microphysics}},\ }\href {https://doi.org/10.1093/mnras/stab1824} {\bibfield
  {journal} {\bibinfo  {journal} {\mnras}\ }\textbf {\bibinfo {volume} {506}},\
  \bibinfo {pages} {3511} (\bibinfo {year} {2021})}\BibitemShut {NoStop}%
\bibitem [{\citenamefont {{Just}}\ \emph {et~al.}(2023)\citenamefont {{Just}},
  \citenamefont {{Vijayan}}, \citenamefont {{Xiong}}, \citenamefont
  {{Bauswein}}, \citenamefont {{Goriely}}, \citenamefont {{Guilet}},
  \citenamefont {{Janka}},\ and\ \citenamefont
  {{Mart{\'\i}nez-Pinedo}}}]{Just2023a}%
  \BibitemOpen
  \bibfield  {author} {\bibinfo {author} {\bibfnamefont {O.}~\bibnamefont
  {{Just}}}, \bibinfo {author} {\bibfnamefont {V.}~\bibnamefont {{Vijayan}}},
  \bibinfo {author} {\bibfnamefont {Z.}~\bibnamefont {{Xiong}}}, \bibinfo
  {author} {\bibfnamefont {A.}~\bibnamefont {{Bauswein}}}, \bibinfo {author}
  {\bibfnamefont {S.}~\bibnamefont {{Goriely}}}, \bibinfo {author}
  {\bibfnamefont {J.}~\bibnamefont {{Guilet}}}, \bibinfo {author}
  {\bibfnamefont {H.-T.}\ \bibnamefont {{Janka}}},\ and\ \bibinfo {author}
  {\bibfnamefont {G.}~\bibnamefont {{Mart{\'\i}nez-Pinedo}}},\ }\bibfield
  {title} {\bibinfo {title} {{End-to-end kilonova models of neutron-star
  mergers with delayed black-hole formation}},\ }\href
  {https://doi.org/10.48550/arXiv.2302.10928} {\bibfield  {journal} {\bibinfo
  {journal} {arXiv e-prints}\ ,\ \bibinfo {eid} {arXiv:2302.10928}} (\bibinfo
  {year} {2023})},\ \Eprint {https://arxiv.org/abs/2302.10928}
  {arXiv:2302.10928 [astro-ph.HE]} \BibitemShut {NoStop}%
\bibitem [{\citenamefont {{Shakura}}\ and\ \citenamefont
  {{Sunyaev}}(1973)}]{shakura1973}%
  \BibitemOpen
  \bibfield  {author} {\bibinfo {author} {\bibfnamefont {N.~I.}\ \bibnamefont
  {{Shakura}}}\ and\ \bibinfo {author} {\bibfnamefont {R.~A.}\ \bibnamefont
  {{Sunyaev}}},\ }\bibfield  {title} {\bibinfo {title} {{Black holes in binary
  systems. Observational appearance.}},\ }\href@noop {} {\bibfield  {journal}
  {\bibinfo  {journal} {A\&A}\ }\textbf {\bibinfo {volume} {24}},\ \bibinfo
  {pages} {337} (\bibinfo {year} {1973})}\BibitemShut {NoStop}%
\bibitem [{\citenamefont {{Stone}}\ \emph {et~al.}(1999)\citenamefont
  {{Stone}}, \citenamefont {{Pringle}},\ and\ \citenamefont
  {{Begelman}}}]{stone1999}%
  \BibitemOpen
  \bibfield  {author} {\bibinfo {author} {\bibfnamefont {J.~M.}\ \bibnamefont
  {{Stone}}}, \bibinfo {author} {\bibfnamefont {J.~E.}\ \bibnamefont
  {{Pringle}}},\ and\ \bibinfo {author} {\bibfnamefont {M.~C.}\ \bibnamefont
  {{Begelman}}},\ }\bibfield  {title} {\bibinfo {title} {{Hydrodynamical
  non-radiative accretion flows in two dimensions}},\ }\href
  {https://doi.org/10.1046/j.1365-8711.1999.03024.x} {\bibfield  {journal}
  {\bibinfo  {journal} {\mnras}\ }\textbf {\bibinfo {volume} {310}},\ \bibinfo
  {pages} {1002} (\bibinfo {year} {1999})},\ \Eprint
  {https://arxiv.org/abs/astro-ph/9908185} {arXiv:astro-ph/9908185 [astro-ph]}
  \BibitemShut {NoStop}%
\bibitem [{\citenamefont {{Lee}}\ \emph {et~al.}(2009)\citenamefont {{Lee}},
  \citenamefont {{Ramirez-Ruiz}},\ and\ \citenamefont
  {{L{\'o}pez-C{\'a}mara}}}]{lee_2009}%
  \BibitemOpen
  \bibfield  {author} {\bibinfo {author} {\bibfnamefont {W.~H.}\ \bibnamefont
  {{Lee}}}, \bibinfo {author} {\bibfnamefont {E.}~\bibnamefont
  {{Ramirez-Ruiz}}},\ and\ \bibinfo {author} {\bibfnamefont {D.}~\bibnamefont
  {{L{\'o}pez-C{\'a}mara}}},\ }\bibfield  {title} {\bibinfo {title} {{Phase
  Transitions and He-Synthesis-Driven Winds in Neutrino Cooled Accretion Disks:
  Prospects for Late Flares in Short Gamma-Ray Bursts}},\ }\href
  {https://doi.org/10.1088/0004-637X/699/2/L93} {\bibfield  {journal} {\bibinfo
   {journal} {\apj}\ }\textbf {\bibinfo {volume} {699}},\ \bibinfo {pages}
  {L93} (\bibinfo {year} {2009})},\ \Eprint {https://arxiv.org/abs/0904.3752}
  {arXiv:0904.3752 [astro-ph.HE]} \BibitemShut {NoStop}%
\bibitem [{\citenamefont {{Seitenzahl}}\ \emph {et~al.}(2008)\citenamefont
  {{Seitenzahl}}, \citenamefont {{Timmes}}, \citenamefont
  {{Marin-Lafl{\`e}che}}, \citenamefont {{Brown}}, \citenamefont
  {{Magkotsios}},\ and\ \citenamefont {{Truran}}}]{seitenzahl_2008}%
  \BibitemOpen
  \bibfield  {author} {\bibinfo {author} {\bibfnamefont {I.~R.}\ \bibnamefont
  {{Seitenzahl}}}, \bibinfo {author} {\bibfnamefont {F.~X.}\ \bibnamefont
  {{Timmes}}}, \bibinfo {author} {\bibfnamefont {A.}~\bibnamefont
  {{Marin-Lafl{\`e}che}}}, \bibinfo {author} {\bibfnamefont {E.}~\bibnamefont
  {{Brown}}}, \bibinfo {author} {\bibfnamefont {G.}~\bibnamefont
  {{Magkotsios}}},\ and\ \bibinfo {author} {\bibfnamefont {J.}~\bibnamefont
  {{Truran}}},\ }\bibfield  {title} {\bibinfo {title} {{Proton-rich Nuclear
  Statistical Equilibrium}},\ }\href {https://doi.org/10.1086/592501}
  {\bibfield  {journal} {\bibinfo  {journal} {\apj}\ }\textbf {\bibinfo
  {volume} {685}},\ \bibinfo {pages} {L129} (\bibinfo {year} {2008})},\ \Eprint
  {https://arxiv.org/abs/0808.2033} {arXiv:0808.2033 [astro-ph]} \BibitemShut
  {NoStop}%
\bibitem [{\citenamefont {Goriely}(1999)}]{Goriely1999}%
  \BibitemOpen
  \bibfield  {author} {\bibinfo {author} {\bibfnamefont {S.}~\bibnamefont
  {Goriely}},\ }\bibfield  {title} {\bibinfo {title} {{Uncertainties in the
  solar system r-abundance distribution}},\ }\href@noop {} {\bibfield
  {journal} {\bibinfo  {journal} {Astron. \& Astrophys.}\ }\textbf {\bibinfo
  {volume} {342}},\ \bibinfo {pages} {881} (\bibinfo {year}
  {1999})}\BibitemShut {NoStop}%
\bibitem [{\citenamefont {{Roederer}}\ \emph {et~al.}(2022)\citenamefont
  {{Roederer}}, \citenamefont {{Lawler}}, \citenamefont {{Den Hartog}},
  \citenamefont {{Placco}}, \citenamefont {{Surman}}, \citenamefont {{Beers}},
  \citenamefont {{Ezzeddine}}, \citenamefont {{Frebel}}, \citenamefont
  {{Hansen}}, \citenamefont {{Hattori}}, \citenamefont {{Holmbeck}},\ and\
  \citenamefont {{Sakari}}}]{Roederer2022}%
  \BibitemOpen
  \bibfield  {author} {\bibinfo {author} {\bibfnamefont {I.~U.}\ \bibnamefont
  {{Roederer}}}, \bibinfo {author} {\bibfnamefont {J.~E.}\ \bibnamefont
  {{Lawler}}}, \bibinfo {author} {\bibfnamefont {E.~A.}\ \bibnamefont {{Den
  Hartog}}}, \bibinfo {author} {\bibfnamefont {V.~M.}\ \bibnamefont
  {{Placco}}}, \bibinfo {author} {\bibfnamefont {R.}~\bibnamefont {{Surman}}},
  \bibinfo {author} {\bibfnamefont {T.~C.}\ \bibnamefont {{Beers}}}, \bibinfo
  {author} {\bibfnamefont {R.}~\bibnamefont {{Ezzeddine}}}, \bibinfo {author}
  {\bibfnamefont {A.}~\bibnamefont {{Frebel}}}, \bibinfo {author}
  {\bibfnamefont {T.~T.}\ \bibnamefont {{Hansen}}}, \bibinfo {author}
  {\bibfnamefont {K.}~\bibnamefont {{Hattori}}}, \bibinfo {author}
  {\bibfnamefont {E.~M.}\ \bibnamefont {{Holmbeck}}},\ and\ \bibinfo {author}
  {\bibfnamefont {C.~M.}\ \bibnamefont {{Sakari}}},\ }\bibfield  {title}
  {\bibinfo {title} {{The R-process Alliance: A Nearly Complete R-process
  Abundance Template Derived from Ultraviolet Spectroscopy of the
  R-process-enhanced Metal-poor Star HD 222925}},\ }\href
  {https://doi.org/10.3847/1538-4365/ac5cbc} {\bibfield  {journal} {\bibinfo
  {journal} {Astrophys. J. Suppl.}\ }\textbf {\bibinfo {volume} {260}},\
  \bibinfo {eid} {27} (\bibinfo {year} {2022})},\ \Eprint
  {https://arxiv.org/abs/2205.03426} {2205.03426} \BibitemShut {NoStop}%
\bibitem [{\citenamefont {{Klion}}\ \emph {et~al.}(2022)\citenamefont
  {{Klion}}, \citenamefont {{Tchekhovskoy}}, \citenamefont {{Kasen}},
  \citenamefont {{Kathirgamaraju}}, \citenamefont {{Quataert}},\ and\
  \citenamefont {{Fern{\'a}ndez}}}]{klion_2022}%
  \BibitemOpen
  \bibfield  {author} {\bibinfo {author} {\bibfnamefont {H.}~\bibnamefont
  {{Klion}}}, \bibinfo {author} {\bibfnamefont {A.}~\bibnamefont
  {{Tchekhovskoy}}}, \bibinfo {author} {\bibfnamefont {D.}~\bibnamefont
  {{Kasen}}}, \bibinfo {author} {\bibfnamefont {A.}~\bibnamefont
  {{Kathirgamaraju}}}, \bibinfo {author} {\bibfnamefont {E.}~\bibnamefont
  {{Quataert}}},\ and\ \bibinfo {author} {\bibfnamefont {R.}~\bibnamefont
  {{Fern{\'a}ndez}}},\ }\bibfield  {title} {\bibinfo {title} {{The impact of
  r-process heating on the dynamics of neutron star merger accretion disc winds
  and their electromagnetic radiation}},\ }\href
  {https://doi.org/10.1093/mnras/stab3583} {\bibfield  {journal} {\bibinfo
  {journal} {\mnras}\ }\textbf {\bibinfo {volume} {510}},\ \bibinfo {pages}
  {2968} (\bibinfo {year} {2022})},\ \Eprint {https://arxiv.org/abs/2108.04251}
  {arXiv:2108.04251 [astro-ph.HE]} \BibitemShut {NoStop}%
\bibitem [{\citenamefont {{Espino}}\ \emph {et~al.}(2023)\citenamefont
  {{Espino}}, \citenamefont {{Bozzola}},\ and\ \citenamefont
  {{Paschalidis}}}]{espino_2022}%
  \BibitemOpen
  \bibfield  {author} {\bibinfo {author} {\bibfnamefont {P.~L.}\ \bibnamefont
  {{Espino}}}, \bibinfo {author} {\bibfnamefont {G.}~\bibnamefont
  {{Bozzola}}},\ and\ \bibinfo {author} {\bibfnamefont {V.}~\bibnamefont
  {{Paschalidis}}},\ }\bibfield  {title} {\bibinfo {title} {{Quantifying
  uncertainties in general relativistic magnetohydrodynamic codes}},\ }\href
  {https://doi.org/10.1103/PhysRevD.107.104059} {\bibfield  {journal} {\bibinfo
   {journal} {\prd}\ }\textbf {\bibinfo {volume} {107}},\ \bibinfo {eid}
  {104059} (\bibinfo {year} {2023})},\ \Eprint
  {https://arxiv.org/abs/2210.13481} {arXiv:2210.13481 [gr-qc]} \BibitemShut
  {NoStop}%
\bibitem [{\citenamefont {{Loken}}\ \emph {et~al.}(2010)\citenamefont
  {{Loken}}, \citenamefont {{Gruner}}, \citenamefont {{Groer}}, \citenamefont
  {{Peltier}}, \citenamefont {{Bunn}}, \citenamefont {{Craig}}, \citenamefont
  {{Henriques}}, \citenamefont {{Dempsey}}, \citenamefont {{Yu}}, \citenamefont
  {{Chen}}, \citenamefont {{Dursi}}, \citenamefont {{Chong}}, \citenamefont
  {{Northrup}}, \citenamefont {{Pinto}}, \citenamefont {{Knecht}},\ and\
  \citenamefont {{Van Zon}}}]{SciNet}%
  \BibitemOpen
  \bibfield  {author} {\bibinfo {author} {\bibfnamefont {C.}~\bibnamefont
  {{Loken}}}, \bibinfo {author} {\bibfnamefont {D.}~\bibnamefont {{Gruner}}},
  \bibinfo {author} {\bibfnamefont {L.}~\bibnamefont {{Groer}}}, \bibinfo
  {author} {\bibfnamefont {R.}~\bibnamefont {{Peltier}}}, \bibinfo {author}
  {\bibfnamefont {N.}~\bibnamefont {{Bunn}}}, \bibinfo {author} {\bibfnamefont
  {M.}~\bibnamefont {{Craig}}}, \bibinfo {author} {\bibfnamefont
  {T.}~\bibnamefont {{Henriques}}}, \bibinfo {author} {\bibfnamefont
  {J.}~\bibnamefont {{Dempsey}}}, \bibinfo {author} {\bibfnamefont {C.-H.}\
  \bibnamefont {{Yu}}}, \bibinfo {author} {\bibfnamefont {J.}~\bibnamefont
  {{Chen}}}, \bibinfo {author} {\bibfnamefont {L.~J.}\ \bibnamefont {{Dursi}}},
  \bibinfo {author} {\bibfnamefont {J.}~\bibnamefont {{Chong}}}, \bibinfo
  {author} {\bibfnamefont {S.}~\bibnamefont {{Northrup}}}, \bibinfo {author}
  {\bibfnamefont {J.}~\bibnamefont {{Pinto}}}, \bibinfo {author} {\bibfnamefont
  {N.}~\bibnamefont {{Knecht}}},\ and\ \bibinfo {author} {\bibfnamefont
  {R.}~\bibnamefont {{Van Zon}}},\ }\bibfield  {title} {\bibinfo {title}
  {{SciNet: Lessons Learned from Building a Power-efficient Top-20 System and
  Data Centre}},\ }in\ \href {https://doi.org/10.1088/1742-6596/256/1/012026}
  {\emph {\bibinfo {booktitle} {Journal of Physics Conference Series}}},\
  \bibinfo {series} {Journal of Physics Conference Series}, Vol.\ \bibinfo
  {volume} {256}\ (\bibinfo {year} {2010})\ p.\ \bibinfo {pages}
  {012026}\BibitemShut {NoStop}%
\bibitem [{\citenamefont {{Ponce}}\ \emph {et~al.}(2019)\citenamefont
  {{Ponce}}, \citenamefont {{van Zon}}, \citenamefont {{Northrup}},
  \citenamefont {{Gruner}}, \citenamefont {{Chen}}, \citenamefont {{Ertinaz}},
  \citenamefont {{Fedoseev}}, \citenamefont {{Groer}}, \citenamefont {{Mao}},
  \citenamefont {{Mundim}}, \citenamefont {{Nolta}}, \citenamefont {{Pinto}},
  \citenamefont {{Saldarriaga}}, \citenamefont {{Slavnic}}, \citenamefont
  {{Spence}}, \citenamefont {{Yu}},\ and\ \citenamefont {{Peltier}}}]{Niagara}%
  \BibitemOpen
  \bibfield  {author} {\bibinfo {author} {\bibfnamefont {M.}~\bibnamefont
  {{Ponce}}}, \bibinfo {author} {\bibfnamefont {R.}~\bibnamefont {{van Zon}}},
  \bibinfo {author} {\bibfnamefont {S.}~\bibnamefont {{Northrup}}}, \bibinfo
  {author} {\bibfnamefont {D.}~\bibnamefont {{Gruner}}}, \bibinfo {author}
  {\bibfnamefont {J.}~\bibnamefont {{Chen}}}, \bibinfo {author} {\bibfnamefont
  {F.}~\bibnamefont {{Ertinaz}}}, \bibinfo {author} {\bibfnamefont
  {A.}~\bibnamefont {{Fedoseev}}}, \bibinfo {author} {\bibfnamefont
  {L.}~\bibnamefont {{Groer}}}, \bibinfo {author} {\bibfnamefont
  {F.}~\bibnamefont {{Mao}}}, \bibinfo {author} {\bibfnamefont {B.~C.}\
  \bibnamefont {{Mundim}}}, \bibinfo {author} {\bibfnamefont {M.}~\bibnamefont
  {{Nolta}}}, \bibinfo {author} {\bibfnamefont {J.}~\bibnamefont {{Pinto}}},
  \bibinfo {author} {\bibfnamefont {M.}~\bibnamefont {{Saldarriaga}}}, \bibinfo
  {author} {\bibfnamefont {V.}~\bibnamefont {{Slavnic}}}, \bibinfo {author}
  {\bibfnamefont {E.}~\bibnamefont {{Spence}}}, \bibinfo {author}
  {\bibfnamefont {C.-H.}\ \bibnamefont {{Yu}}},\ and\ \bibinfo {author}
  {\bibfnamefont {W.~R.}\ \bibnamefont {{Peltier}}},\ }\bibfield  {title}
  {\bibinfo {title} {{Deploying a Top-100 Supercomputer for Large Parallel
  Workloads: the Niagara Supercomputer}},\ }\href@noop {} {\bibfield  {journal}
  {\bibinfo  {journal} {arXiv e-prints}\ ,\ \bibinfo {eid} {arXiv:1907.13600}}
  (\bibinfo {year} {2019})},\ \Eprint {https://arxiv.org/abs/1907.13600}
  {arXiv:1907.13600 [cs.DC]} \BibitemShut {NoStop}%
\end{thebibliography}%

\end{document}